\documentclass[english,onecolumn]{IEEEtran}
\usepackage[T1]{fontenc}
\usepackage{babel}
\usepackage{array}
\usepackage{float}
\usepackage{amsmath}
\usepackage{amsthm}
\usepackage{amssymb}
\usepackage{graphicx}
\usepackage{setspace}
\usepackage[unicode=true,
 bookmarks=true,bookmarksnumbered=true,bookmarksopen=true,bookmarksopenlevel=1,
 breaklinks=false,pdfborder={0 0 0},pdfborderstyle={},backref=false,colorlinks=false]
 {hyperref}
\hypersetup{pdftitle={Your Title},
 pdfauthor={Your Name},
 pdfpagelayout=OneColumn, pdfnewwindow=true, pdfstartview=XYZ, plainpages=false}

\makeatletter

\providecommand{\tabularnewline}{\\}
\floatstyle{ruled}
\newfloat{algorithm}{tbp}{loa}
\providecommand{\algorithmname}{Algorithm}
\floatname{algorithm}{\protect\algorithmname}

\let\oldforeign@language\foreign@language
\DeclareRobustCommand{\foreign@language}[1]{%
  \lowercase{\oldforeign@language{#1}}}
\theoremstyle{plain}
\newtheorem{thm}{\protect\theoremname}[section]
\theoremstyle{definition}
\newtheorem{defn}[thm]{\protect\definitionname}
\theoremstyle{plain}
\newtheorem{prop}[thm]{\protect\propositionname}
\theoremstyle{plain}
\newtheorem{lem}[thm]{\protect\lemmaname}
\theoremstyle{plain}
\newtheorem{cor}[thm]{\protect\corollaryname}
\theoremstyle{remark}
\newtheorem{rem}[thm]{\protect\remarkname}

\usepackage[figurename=Fig.]{caption}
\usepackage[caption=false,font=footnotesize]{subfig}
\usepackage{cite}
\DeclareMathOperator*{\plim}{plim}

\makeatother

\providecommand{\corollaryname}{Corollary}
\providecommand{\definitionname}{Definition}
\providecommand{\lemmaname}{Lemma}
\providecommand{\propositionname}{Proposition}
\providecommand{\remarkname}{Remark}
\providecommand{\theoremname}{Theorem}

\begin{document}

\title{Game Theoretic Dynamic Channel Allocation for Frequency-Selective
Interference Channels}

\author{Ilai~Bistritz,~\IEEEmembership{Student Member,~IEEE} , Amir~Leshem,~\IEEEmembership{Senior Member,~IEEE}
\thanks{Parts of this paper were presented at the 53th Annual Allerton Conference
on Communication, Control, and Computing, Monticello, IL, September
2015 \cite{Bistritz2015} and in the 42nd IEEE International Conference
on Acoustics, Speech and Signal Processing \cite{Bistritz2017}.\protect \\
Ilai Bistritz is with the department of Electrical Engineering-Systems,
Tel-Aviv University, Israel, e-mail: ilaibist@gmail.com.\protect \\
Amir Leshem~is with the Faculty of Engineering, Bar-Ilan University,
Ramat-Gan, Israel, e-mail: leshema@eng.biu.ac.il.

This research was supported by the Israel Science Foundation, under
grant 2277/2017, and partially supported by the Israeli Ministry of
Science and Technology under grant 3-13038.}}

\markboth{}{Your Name \MakeLowercase{\emph{et al.}}: Your Title}

\IEEEpubid{}
\maketitle
\begin{abstract}
We consider the problem of distributed channel allocation in large
networks under the frequency-selective interference channel. Performance
is measured by the weighted sum of achievable rates. Our proposed
algorithm is a modified Fictitious Play algorithm that can be implemented
distributedly and its stable points are the pure Nash equilibria of
a given game. Our goal is to design a utility function for a non-cooperative
game such that all of its pure Nash equilibria have close to optimal
global performance. This will make the algorithm close to optimal
while requiring no communication between users. We propose a novel
technique to analyze the Nash equilibria of a random interference
game, determined by the random channel gains. Our analysis is asymptotic
in the number of users. First we present a natural non-cooperative
game where the utility of each user is his achievable rate. It is
shown that, asymptotically in the number of users and for strong enough
interference, this game exhibits many bad equilibria. Then we propose
a novel non-cooperative M Frequency-Selective Interference Channel
Game (M-FSIG), as a slight modification of the former, where the utility
of each user is artificially limited. We prove that even its worst
equilibrium has asymptotically optimal weighted sum-rate for any interference
regime and even for correlated channels. This is based on an order
statistics analysis of the fading channels that is valid for a broad
class of fading distributions (including Rayleigh, Rician, m-Nakagami
and more). We carry out simulations that show fast convergence of
our algorithm to the proven asymptotically optimal pure Nash equilibria.
\end{abstract}

\begin{IEEEkeywords}
Frequency-selective fading channels, ad hoc networks, resource allocation,
random games, utility design.
\end{IEEEkeywords}

\IEEEpeerreviewmaketitle{}

\section{Introduction}

\IEEEPARstart{T}{he} problem of allocating bandwidth to users is
a fundamental component in every wireless network. The scenario of
users that share a common wireless medium is known as the interference
channel. The capacity region of the interference channel is not yet
exactly known \cite{TeSun1981}, and even for the two-user Gaussian
flat channel case it is only known to within one bit \cite{Etkin2008}.
Sequential cancellation techniques, which lead to the best known achievable
rate region in the strong-interference case, are impractical for large
networks. Interference alignment techniques \cite{Cadambe2008,ElAyach2013}
can theoretically achieve half the degrees of freedom in any interference
regime but require each user to have all the channel state information
(CSI) of all users.

Obtaining all the necessary CSI in large networks is a fundamental
issue that might require a great deal of communication between the
users. This communication requirement is somewhat paradoxical since
the process of channel allocation, by its nature, should be already
established for communication between users to be even possible. Furthermore,
in a wireless environment the channel gains are time varying and can
be assumed constant only within the coherence time of the channel.
Gathering all the channel gains in a large network might take more
time than the coherence time - hence renders itself useless. Thus,
one can ask if the traditional capacity region successfully measures
the potential utilization of the interference channel in distributed
scenarios in large networks.

When coordination is infeasible, how can we identify achievable rate
points? In such a scenario, users act selfishly, not out of viciousness
but out of ignorance. Each user simply does not hold enough information
about other users for considering their interests in his decision.
This makes each user an independent decision maker, which naturally
calls for a game-theoretic analysis. The interplay between game theory
and information theory has attracted a lot of attention from researchers
in the last decade \cite{Leshem2006,Leshem2008,Laufer2005,Pang2008,Berry2011,Noam2009,Jorswieck2013,Alpcan2009,Gao2014}.
In this work we focus on the implications of this interplay for the
interference channel. We provide and employ two key observations.

The first observation is that if a certain Nash equilibrium (NE) exists
it does not mean that this NE is also achievable. Even if good NE
do exist, the dynamics do not necessarily converge to a good NE when
poor NE also exist. The problem of tuning the dynamics to a specific
equilibrium (equilibrium selection) is generally difficult and may
require some coordination between users. Therefore, suggestions have
been made to measure the cost of this uncertainty about the resulting
NE by the price of anarchy \cite{Koutsoupias1999}, or equivalently
by the performance of the worst equilibrium.

The second observation is that the utility function is a design parameter.
It can be chosen to optimize the performance of the NE. Examples for
this approach include pricing mechanisms \cite{Saraydar2002,Alpcan2009,Alpcan2009a}
and utility design \cite{Marden2013,Gopalakrishnan2011}. The users
by no means have free will. They use machines (transceivers) that
act according to a predefined program or protocol. A utility function
simply defines the user's decision rule (his device's program). Every
utility function yields an admissible decision rule as long as it
can be computed fully distributedly and maximized by each user independently.

In this work, we combine the above two observations to create a ``min-max''
argument - in a fully distributed scenario, the achievable NE are
the worst NE of the best game (i.e., with the best utility function).
We demonstrate this approach for the case of the frequency-selective
interference channel (see \cite{Larsson2009,Scutari2008,Leshem2009}).
The simplest and best known achievable points in the interference
channel are those that result from orthogonal transmission schemes
such as TDMA/FDMA or CDMA, and the more recent bandwidth efficient
techniques such as OFDMA and SC-FDMA \cite{Tsiropoulou2016}. This
is also the most common way to access the channel in practice. In
the spirit of these schemes, we consider a scenario with $N$ users
and $N$ resource elements (REs), where a single RE is a time-frequency
slot. We show that stable point (NE) channel allocations with asymptotically
optimal weighted sum-rate are achievable fully distributedly. These
NE naturally implement frequency-reuse (resource sharing) for users
who are spatially separated. Note that if the allocation is not completely
orthogonal then we assume that the resulting interference is treated
as noise.

Other competitive approaches based on iterative water filling (IWF,
see \cite{Yu2002,Scutari2008a,Scutari2009}) allow users to allocate
their power over the spectrum as a whole. According to \cite{Zhang2015},
a distributed water filling scheme might result in NE with sum-rate
performance loss as compared to the NE of a single channel selection
scheme\@. This result puts in question the use of distributed power
allocation algorithms, in favor of allowing each user to select a
single channel.

In the interference channel, the maximum achievable sum-rate for orthogonal
transmission schemes is the capacity of the single user in the original
channel (assuming identical transmission powers). In the frequency-selective
channel, different users experience different conditions in each channel
due to fading in addition to interference, so different allocations
will result in varying levels of performance. Assigning each user
a good channel results in a gain known as multi-user diversity \cite{Knopp1995}.
In fact, if the allocation method is fast enough, users can maintain
good frequency bands and the network throughput can be significantly
increased.

\subsection{Contributions of this paper}

In this paper we both suggest a novel distributed channel allocation
algorithm based on utility design, and provide a novel analysis for
NE in random games.

\subsubsection{Random Game NE analysis}

The vast majority of the existing literature in game theory focuses
on analyzing fixed games with fixed utility functions and fixed parameters.
There are some famous results for the existence of pure NE in special
cases of games such as potential games, supermodular games, games
with quasi-concave utilities and more (see \cite{Han2012}).

As opposed to fixed games, in Bayesian games (see \cite{Han2012})
there is a distribution for the random parameters of the game. A Bayesian
NE is computed with respect to the expected utility over these random
parameters. This can be thought of as the NE of the ``average game''.
Considering the average game instead of the actual one that is going
to happen can lead to weak results. This is enhanced when one is interested
in games that converge to a solution that has a good global performance
guarantee. It would have been better to analyze instead the distribution
of the random NE of the random game.

The NE points of our game are determined by the channel gains, which
we model as random. This leads us to analyze a random game. To the
best of our knowledge, little work has been done on random games.
From a game-theoretic point of view, various issues related to random
games have been addressed, such as the number of pure Nash equilibria
and the complexity of finding a Nash equilibrium (see \cite{Barany2007,Rinott2000,Cohen1998,Daskalakis2011}).
The common model for a random game assumes that the payoff vectors
are i.i.d. for different strategy profiles. This assumption can be
interpreted as a lack of structure for the random game. Our approach
is essentially different, as our game is chosen at random from all
the games with some structure of interest. This structure stems directly
from the physical reality; namely, the wireless environment in our
case.

For interference networks, the only existing works that analyzed the
NE of random networks did so for potential games \cite{Perlaza2013}
or two-player games \cite{Rose2001}. In a potential game, at least
one pure NE is guaranteed to exist, and this pure NE is the maximum
of the potential function. This means that the pure NE can be expressed
in a closed-form as a function of the random channel gains in the
network. In this case there is no random NE existence analysis. This
should not be confused with the fact that the performance of this
known NE is a random variable, and can be analyzed as one.

In this work we introduce a novel technique for random NE existence
analysis in random games. We analyze the random structure of these
random NE and exploit the large number of users to provide concentration
results on this random structure. In our case the random structure
of the NE is of ``almost'' a perfect matching in a random bipartite
graph. However, this technique can be applied to other game-theoretic
problems, beyond the case of channel allocation, by identifying the
random structure of the NE.

In our channel allocation scenario the number of users is proportional
to the number of (frequency) channels. If the large number of users
were to be much larger than the number of channels, the average number
of users per channel would become large. In this case concentration
results can be applied directly to the total interference in each
channel. This allows for greatly simplifying the NE analysis of the
game. For example, one could use Congestion games, Wardrop equilibrium
or Evolutionary and Mean-Field games (see \cite{Han2012}). Our work
contributes to the set of tools offered by the limit of large networks
in a scenario where each individual user does not become negligible
or anonymous as the number of users approaches infinity.

\subsubsection{Utility Design for Distributed Channel Allocation}

Our work is the first to introduce a fully distributed channel allocation
algorithm for the frequency-selective channel, that requires no communication
between users and still achieves a close to optimal sum-rate performance
(while maintaining fairness) in all of its equilibria, for large networks.
We do so by constructing a non-cooperative game between the users.
In contrast to many game-theoretic works, \textbf{we do not view the
utility function as a model for what rational users might want to
optimize, but as a design choice that the devices are programmed to
maximize distributedly}. Hence, our game induces a novel MAC protocol.
This game-theoretic approach is generally known as utility design.

Utility design is the process of choosing a utility function for a
non-cooperative game such that its NE will exhibit good global performance.
Utility design is a very attractive distributed optimization tool,
since it tends to yield algorithms that require no coordination between
users, have very flexible synchronization requirements and may be
applied to non-standard global objectives (e.g., non-convex objectives).
For comparison, the wide-spread network utility maximization (NUM,
see \cite{Kelly1998,Palomar2006}) approach requires communication
between users to be able to distribute the computation, and is generally
applicable only when the objective is a sum of convex functions.

Utility design is similar in spirit to mechanism design \cite{Han2012}.
However, mechanism design deals with users that are not programmed
agents but strategic users who can deliberately manipulate the system.
Furthermore, mechanism design traditionally requires a game manager,
while no central entity of any sort exists in our scenario.

The inherent difficulty in utility design is the vast, or simply undefined,
optimization domain - all the options for a utility function. One
approach is to limit the utility function to a certain form with parameters,
and optimize the performance of the resulting NE over the domain of
these design parameters. This approach appears in pricing mechanisms,
where a linear term is subtracted from the utility function, called
the price \cite{Alpcan2009,Saraydar2002,Alpcan2009a}. By its nature,
the pricing approach is only applicable to games with continuous strategy
spaces - which is not our case.

Utility design techniques for discrete resource allocation scenarios
are presented in \cite{Marden2013,Gopalakrishnan2011}. Some of the
utility designs in these works result in a potential game with a price
of anarchy of two. These designs are based on the marginal contributions
of each user to the global performance function (social-welfare).
Unfortunately, in an interference network, it is highly unreasonable
to require from each user to know his marginal contributions, since
they depend on the channel gains of other users.

Our approach achieves a price of anarchy close to one (instead of
two) using a utility that does not require computing the marginal
contributions but only requires locally and naturally available information.
Our design starts by first analyzing the selfish utility, where each
user maximizes his achievable rate. Then, based on the understanding
of why this choice failed, we propose a slight modification of the
selfish utility that turns out to be the optimal design, at least
asymptotically in the number of users.

There exist many distributed channel allocation schemes there are
not based on game-theoretic approaches. It has been shown that under
less restrictive demands, the optimal solution to the resource allocation
problem can be achieved using a distributed algorithm \cite{Naparstek2013,Naparstek2014,Naparstek2014a},
but with a very slow convergence rate. In \cite{Leshem2012} a distributed
algorithm based on the stable matching concept has been proposed.
This algorithm has a much faster convergence rate than the previous
one and a good sum-rate performance, but not necessarily close to
the optimal. In \cite{Cohen2013} a distributed ALOHA based algorithm
was shown to converge to a NE with good (but not optimal) performance.
The main disadvantage of all of these algorithms is their assumption
that each user, without exceptions, can sense the transmission of
each other user. Maintaining this assumption requires a central network
entity or at least communication between users, thus negatively impacts
the network distributed nature and its scalability (e.g. an RTS/CTS
mechanism for the CSMA based algorithms). It also adds extra delays
to the network.

In our algorithm users choose channels that are good for them without
considering any other users. Hence, no assumption on which user can
sense which other user is needed. Despite this seemingly competitive
algorithm, close to optimal performance, that maintains fairness,
is obtained. Another issue is the synchronization between users these
algorithms require, that is naturally avoided by a game-theoretic
algorithm. In our algorithm, users only need to have their own channel
state information, to be able to measure the interference in each
channel and to have a feedback channel from the receiver to the transmitter.
These three capabilities are very common in modern communication networks.

Additionally, many of the existing works on channel allocation for
the frequency-selective channel only consider the case of Rayleigh
fading with i.i.d. channel gains (see \cite{Naparstek2013,Naparstek2014,Naparstek2014a,Perlaza2013,Rose2001,Zhang2015}).
Our work offers a general analysis that is valid for both a broad
class of fading distributions and correlated channel gains.

\subsection{Outline}

The rest of this paper is organized as follows. In Section II we formulate
our wireless network model and define our global objective function.
In Section III we present our game-theoretic approach, which is to
analyze the probability of NE of a random interference game asymptotically
in the number of users. Our algorithm is a modified fictitious play
algorithm that converges to these NE and can be implemented fully
distributedly by each user. In Section IV we present a natural game
formulation for this problem, where users maximize their own achievable
rate. We prove in Theorem \ref{NFSIG-PPOA} that it may lead to poor
performance with strong enough interference. In Section V we propose
instead a game with a carefully designed utility function, which is
but a slight modification of the former. We prove in Theorem  \ref{M-FSIG Theorem}
that this utility choice guarantees that all pure NE are asymptotically
optimal for a broad class of fading distributions. In Section VI we
present simulations of our proposed algorithm that show fast convergence
to the proven equilibria and suggest that our asymptotic analysis
is already valid for small values of the number of users and channels.
Finally, we draw conclusions in Section VII.

\section{System Model}

Consider a wireless network consisting of $N$ transmitter-receiver
pairs (users) and $K$ resource elements (REs). Each user forms a
link between his transmitter and receiver using a single resource
element. Throughout this paper we assume for simplicity that $N=K$.
However, the number of frequency channels can be smaller than $N$
by employing also a time division (such as in FDMA/TDMA or OFDMA).
Hence, in general, each RE is a time-frequency slot. The pairs are
formed from a distributed multi-hop routing mechanism \cite{Chang2000,Chang2004}
that we assume arbitrary and is outside the scope of this paper. The
pairing may be constant in the case of a two-way radios networks,
or require a new allocation each time the destinations change. The
users are located in a certain geographical area and some of them
may be far away from each other. Our resulting allocation resembles
an orthogonal transmission. However, it allows for distributedly exploiting
the geometry of the network by allocating two far-away users the same
RE (i.e., frequency-reuse). This models a general ad-hoc network with
no infrastructure, such as sensor networks, device-to-device (D2D)
and cognitive radio networks.

Our goal is to design a fully distributed MAC protocol for channel
allocation with close to optimal performance. We adopt a game-theoretic
approach for this purpose. The structure of our protocol is described
in Algorithm \ref{alg:Modified-Fictitious-Play}. In Section III.3
we prove that the stable points of Algorithm \ref{alg:Modified-Fictitious-Play}
are the pure NE of the designed game. Most of this work is dedicated
to characterize the random PNE of two different utility designs and
analyze their performance. We start from a naive utility choice in
Section IV and modify it in Section V to an asymptotically optimal
design.

As any other MAC protocol, Algorithm \ref{alg:Modified-Fictitious-Play}
is wired in the device of each user and controls how he accesses the
medium. This includes the fact that the protocol employs a non-cooperative
game with a designed utility. Therefore, the game is not played by
users with selfish interests but between devices that were programmed
to implement Algorithm \ref{alg:Modified-Fictitious-Play}, with the
utility function as a design parameter.

Note that Steps b,c,d are all carried at the receiver of user $n$.
In Step a, the receiver of user $n$ has to feedback to transmitter
$n$ the index of the chosen RE. This can be done using a narrow control
channel using a very short packet of $\log_{2}N$ bits, which is further
reduced to $O\left(\log\left(\log N\right)\right)$ bits using the
utility of Section V. This is the only form of feedback required in
the link of user $n$.

\begin{figure}[tbh]
~~~~~~~~~~~~~~~~~~~~~~~~~~~~~~~~~~~~~~~~~~~~~~~~~\includegraphics[width=6cm,height=6cm]{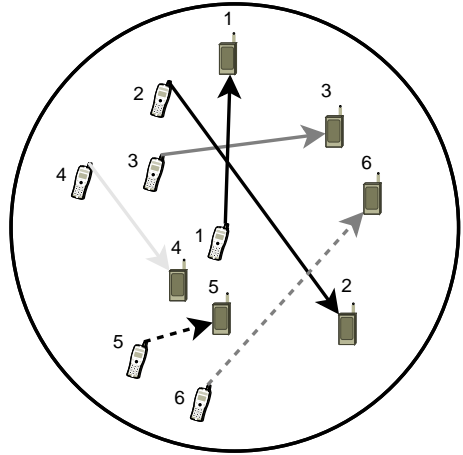}

\caption{\label{fig:System-Model}System Model}
\end{figure}

A toy example of our system is depicted in Fig. \ref{fig:System-Model}.
Transmitters are represented by white devices and receivers by grey
devices. Different types of lines represent links that use different
REs. In this example, User 1 and User 2 use the same RE. This represents
a case where the receivers are far away, and due to their multipath,
both perceive the same RE as better than the unoccupied sixth RE.

\begin{algorithm}[tbh]
\caption{\label{alg:Modified-Fictitious-Play}Game-Theoretic Distributed\textbf{
}Channel\textbf{ }Allocation}

\textbf{Parameters}
\begin{enumerate}
\item Let $h_{m,n,k}$ be the channel gain of RE $k$ (time-frequency slot)
between transmitter $m$ and receiver $n$.
\item Let $P_{n}$ be the transmission power of transmitter $n$.
\item Let $u_{n}\left(\boldsymbol{\mathbf{a}}\right)$ be the utility function
of an FSIG $G$ (see Definition \ref{def:FSIG}).
\end{enumerate}
\textbf{Modified Fictitious Play (Section III.3)}
\begin{enumerate}
\item \textbf{Initialization} - Choose $0<\alpha\leq1$ and $\tau>0$. Each
user picks a RE at random and then initializes his fictitious utility
- $\bar{U}_{n,k}(0)=u_{n}\left(k,\boldsymbol{\mathbf{a}}_{-n}\left(0\right)\right)$
for each $k$.
\item \textbf{At each turn $t>0$, each of the users $n=1,...,N$ does the
following}

\begin{enumerate}
\item Chooses a transmission RE
\[
a_{n}(t)=\arg\underset{k}{\max}\,\overline{U}_{n,k}(t-1)
\]
\item Senses the interference at each RE $k$
\[
I_{k}(\boldsymbol{a}_{-n}\left(t\right))=\underset{m|a_{m}(t)=k}{\sum}|h_{m,n,k}|^{2}P_{m}
\]
\item Updates the fictitious utility for each RE $k$
\[
\overline{U}_{n,k}(t)=(1-\alpha)\overline{U}_{n,k}(t-1)+\alpha u_{n}\left(k,I_{k}\left(\boldsymbol{\mathbf{a}}_{-n}\left(t\right)\right)\right)
\]
\item (optional) Checks for convergence to a PNE of $G$. If $t\mod\tau=0$
and $I_{k}\left(\boldsymbol{\mathbf{a}}_{-n}\left(t\right)\right))\neq I_{k}\left(\boldsymbol{\mathbf{a}}_{-n}\left(t-1\right)\right))$
for some $k$ then set $\bar{U}_{n,k}(t)=0$ for each $k$.
\end{enumerate}
\end{enumerate}
\end{algorithm}

\subsection{Channel Gain Assumptions}

The channel between each transmitter and receiver is Gaussian frequency-selective.
This models an environment where the multipath phenomenon creates
a superposition of waves that makes some frequency bands better than
others, independently between users. The channel gains (fading coefficients)
are modeled as $N^{2}K$ random variables - one for each RE, each
transmitter and each receiver. These channel gains incorporate both
the geometry of the network (in their expectation) and multipath affects
(in their realization). Each of the $K$ REs is a time-frequency slot.
We denote the total number of time slots by $s$ and the total number
of frequency bands by $b$, so $K=bs$. The channel gain between user
$n_{1}$'s transmitter and user $n_{2}$'s receiver in RE $k$ is
denoted $h_{n_{1},n_{2},k}$, for $k=1,...,K$. Without loss of generality,
we assume that all the REs of the same frequency-band have consecutive
indices. For example, $h_{n_{1},n_{2},1}$,..., $h_{n_{1},n_{2},s}$
have a common frequency band and are identical.

While assuming i.i.d. channel gains may facilitate the analysis, in
practice the channel gains may not be statistically independent. Correlations
between channel gains of different REs can occur if the REs have a
common frequency band, or even have frequency bands with carrier frequencies
that are closer than the coherence bandwidth of the channel. For a
non-vanishing transmission rate, the bandwidth of a single frequency
band must be non-decreasing with $N$. Hence, the coherence bandwidth
of the channel is at most $d$ times larger than the frequency band,
for some integer $d$ that is fixed with respect to $N$. For that
reason, with $s$ time slots, only frequency bands with index $j$
such that $\left|i-j\right|\leq ds$ are correlated with the $i$-th
frequency band. Both the time-division and the frequency-division
correlation are well captured by the following notion.
\begin{defn}
A random process $\left\{ X_{i}\right\} $ is said to be $m-$dependent
if and only if for each $i,j$ such that $\left|i-j\right|>m$ the
variables $X_{i}$ and $X_{j}$ are statistically independent.
\end{defn}
Our assumptions regrading the channel gains are defined as follows.
\begin{defn}
\label{def:Channel Gain Assumptions}Let $m$ be a non-negative integer.
The channel gains $\left\{ h_{n_{1},n_{2},k}\right\} $ are said to
form an $m$-dependent frequency-selective channel if all the following
conditions hold
\begin{enumerate}
\item For each $n_{1},n_{2}$, the variables $h_{n_{1},n_{2},1},\,...,\,h_{n_{1},n_{2},K}$
are identically distributed, where $h_{n_{1},n_{2},k}$ is a complex
number and $\left|h_{n_{1},n_{2},k}\right|$ has zero probability
to be zero.
\item For each $k=1,...,K$, the variables $h_{n_{1},n_{2},k}$ and $h_{n_{3},n_{4},k}$
are independent if $n_{1}\neq n_{3}$ or $n_{2}\neq n_{4}$.
\item For each $n_{1},n_{2}$, the variables $h_{n_{1},n_{2},1},\,...,\,h_{n_{1},n_{2},K}$
are $m$-dependent, i.e., for each $i,j$ such that $\left|i-j\right|>m$,
the variables $h_{n_{1},n_{2},i}$ and $h_{n_{1},n_{2},j}$ are independent.
\end{enumerate}
Specifically, we use the term independent frequency-selective channel
instead of an $0$-dependent frequency-selective channel, since $h_{n_{1},n_{2},1},\,...,\,h_{n_{1},n_{2},K}$
are simply independent in this case.
\end{defn}
The independence of channel gains of different users (Part 2 of Definition
\ref{def:Channel Gain Assumptions}) stems from their different physical
locations (see Fig. \ref{fig:System-Model}), where the multipath
pattern is different. As in MIMO transceivers, a distance of half
a wavelength is far enough for the channel gains to be considered
independent. For different users that typically are meters apart,
this independence assumption is a common practice.

Note that $NK$ of the channel gains are between a transmitter and
his designated receiver (a link). These channel gains are referred
to as the direct channel gains. The other $NK(N-1)$ channel gains
serve as interference coefficients between transmitters and unintended
receivers. In a distributed ad-hoc network, we expect each user to
have knowledge only on the channel gains he can measure, as formalized
next.
\begin{defn}[CSI Knowledge Assumptions]
\label{def:CSI Assumptions}We assume that each user $n$ has perfect
channel state information (CSI) of all his $K$ channel gains, $h_{n,n,1},...,h_{n,n,K}$.
We also assume that each user knows the interference $I_{k}\left(\mathbf{a}_{-n}\right)=\underset{m|a_{m}=k}{\sum}|h_{m,n,k}|^{2}P_{m}$
he experiences in each RE $k$, where $P_{m}$ is the transmission
power of user $m$. Users do not have any knowledge regarding the
channel gains of other users and the specific interference coefficients.
\end{defn}
In practice, user $n$ (transmitter-receiver pair) can estimate $h_{n,n,1},...,h_{n,n,K}$
by transmitting a wideband pilot sequence over the channels. The measurement
of the aggregated interference is done in the receiver by directly
measuring the incoming power at each frequency band. This requires
from each device to have a wide-sensing capabilities, that are already
practical in modern devices.

Mathematically, our channel gains are static random variables and
not time-varying random processes. The implication of this model is
that channel gains are constant with time. In practice, channel gains
can only be assumed constant for a duration known as the coherence
time of the channel. In a centralized system, acquiring the large
number of channel gains is likely to take more time than the coherence
time. Our fully distributed approach does not suffer from this limitation.
However, for our channel allocation to be practical, we need an algorithm
that converges to an allocation fast enough. If it converges much
faster than the coherence time of the channel, it can be repeatedly
updated in the network and maintain multi-user diversity. Although
dynamics and convergence are not the emphasis of this work, we do
show in simulations that Algorithm \ref{alg:Modified-Fictitious-Play}
converges very fast. This suggests that our algorithm can be considered
practical in quasi-static environments like sensor networks and in-door
networks, or in general in underspread systems which constitute a
significant part of modern OFDM systems.

\subsection{Performance}

Somewhat surprisingly, we show in this paper that asymptotically optimal
performance can be achieved without the knowledge of the whole CSI
in any part of the network. In fact, our algorithm does not require
any communication at all between users. Our global performance metric
is the weighted sum of achievable rates while treating interference
as noise. We assume the following on the weights.
\begin{defn}[Bounded Weights]
\label{def:Bounded Weights}We assume that the weights satisfy $w_{min}\leq w_{n}\leq w_{max}$
for some $w_{min},w_{max}>0$, for all $n$. Throughout this paper
we refer to this assumption as ``bounded weights''.
\end{defn}
Our global performance metric is defined as follows.
\begin{defn}
\label{def:Global Performance}Denote by $\mathbf{a}$ the allocation
vector (the strategy profile), such that $a_{n}=k$ if user $n$ is
using RE $k$. We want to maximize the following global performance
function over all possible allocations
\begin{equation}
W\left(\mathbf{a}\right)=\sum_{n=1}^{N}w_{n}R_{n}\left(\mathbf{a}\right)=\sum_{n=1}^{N}w_{n}\log_{2}\left(1+\frac{P_{n}|h_{n,n,a_{n}}|^{2}}{N_{0}+I_{a_{n}}\left(\mathbf{a}_{-n}\right)}\right)\label{eq:1-1}
\end{equation}
where $R_{n}\left(\mathbf{a}\right)$ is the achievable rate of user
$n$, $N_{0}$ is the Gaussian noise variance in each receiver, $P_{n}$
is user $n$'s transmission power and $I_{k}\left(\mathbf{a}_{-n}\right)=\underset{m|a_{m}=k}{\sum}|h_{m,n,k}|^{2}P_{m}$
is the interference user $n$ experiences in RE $k$.
\end{defn}
Each user first needs to measure his experienced interference $I_{a_{n}}\left(\mathbf{a}_{-n}\right)$
to be able to compute the achievable rate in \eqref{eq:1-1} and devise
his coding scheme accordingly. This can be done using a short preamble
of pilots sent in the beginning of the transmission. Alternatively,
the transmission scheme can be adapted using a fast enough feedback
in the user's link based on the bit-error rate experienced in the
receiver.

Since our global performance metric includes the weights $\left\{ w_{n}\right\} $,
one would expect that they will appear as an input for the algorithm.
This might be a tricky thing to do in a distributed algorithm, because
it is not clear what is exactly the input and which user should know
which weights. Fortunately, this dilemma is avoided in our case, and
for a very satisfying reason. Our NE of the designed game have asymptotically
optimal weighted sum-rate in $N$, regardless of the choice of weights
as long as they are bounded (independent of $N$). This means that
the NE maximize the sum-rate while maintaining some fairness between
users. Only an asymptotically negligible amount of users might suffer
from not close to optimal performance.

\section{Frequency-Selective Interference Games}

We want to find a fully distributed way to achieve close to optimal
solutions for our channel allocation problem. Hence we need to analyze
the interaction that results from $N$ independent decision makers
and ensure that the outcome is desirable. The natural way to tackle
this problem is by applying game theory. A good overview of game theory
can be found in \cite{Owen1995}. The book by \cite{Han2012} provides
a good overview of applications of game theory to communication. The
two games we analyze in this work belong to the following family.
\begin{defn}
\label{def:FSIG}A frequency-selective interference game (FSIG) is
a normal-form game
\begin{equation}
G=<\mathcal{N},\{A_{n}\}_{n\in\mathcal{N}},\{u_{n}\}_{n\in\mathcal{N}}>\label{eq:2-1}
\end{equation}
 where $\mathcal{N}$ is the set of users, which are transmitter-receiver
pairs (links), $A_{n}$ is the set of the $K$ REs for all $n$, and
$u_{n}:A_{1}\times...\times A_{N}\rightarrow\mathbb{R}$ is the utility
function of user $n$, which is in the form
\begin{equation}
u_{n}\left(k,\boldsymbol{\mathbf{a}}_{-n}\right)=u_{n}\left(k,I_{k}\left(\boldsymbol{\mathbf{a}}_{-n}\right)\right)\label{eq:2}
\end{equation}
where $I_{k}\left(\mathbf{a}_{-n}\right)$ is the interference user
$n$ measures at RE $k$.
\end{defn}
The utility form in \eqref{eq:2} ensures that the utility can be
computed independently by each user by measuring the interference
in each RE (see the CSI Knowledge assumptions of Definition \ref{def:CSI Assumptions}).
This is necessary in order to use $G$ for the distributed MAC protocol
in Algorithm \ref{alg:Modified-Fictitious-Play}.

In the next subsection, we show that the stable points of Algorithm
\ref{alg:Modified-Fictitious-Play} are the pure NE of the FSIG the
algorithm uses.
\begin{defn}
A strategy profile $(a_{n}^{*},\mathbf{a}_{-n}^{*})\in A_{1}\times...\times A_{N}$
is called a pure Nash equilibrium (PNE) if $u_{n}(a_{n}^{*},\mathbf{a}_{-n}^{*})\geq u_{n}(a_{n},\mathbf{a}_{-n}^{*})$
for all $a_{n}\in A_{n}$ and all $n\in\mathcal{N}$.
\end{defn}
This means that for each player $n$, if the other players act according
to the PNE, player $n$ can not improve his utility with another strategy.
A game may exhibit a unique PNE, multiple PNE or none at all.

The utility functions are a design choice that we wish would induce
a FSIG with only good NE in terms of the global performance of Definition
\ref{def:Global Performance}. The measure of success of the utility
choice is given by the Pure Price of Anarchy (PPoA, see \cite{Koutsoupias1999}),
defined as follows.
\begin{defn}
\label{PPoA Definition}The pure price of anarchy (PPoA) of a game
$G=<\mathcal{N},\{A_{n}\}_{n\in\mathcal{N}},\{u_{n}\}_{n\in\mathcal{N}}>$
with the global performance function $W:A_{1}\times...\times A_{N}\rightarrow\mathbb{R}$
is $\frac{\underset{\mathbf{a}\in A_{1}\times...\times A_{N}}{\max}W(\mathbf{a})}{\underset{\mathbf{a}\in\mathcal{P}_{e}}{\min}W(\mathbf{a})}$,
where $\mathcal{P}_{e}$ is the set of PNE.
\end{defn}
A PPoA close to one means that the worst PNE of the game is close
to the optimal solution. This is of great value since guaranteeing
that the dynamics will avoid the worst equilibrium can be extremely
hard, especially without any communication between the users (which
we forbid in this work). Note that the PPoA is defined with respect
to the global performance function (in our case given by Definition
\ref{def:Global Performance}) and not the sum of utilities. The choice
of the utility function affects the PPoA only indirectly, by determining
the PNE of the resulting game, and specifically the worst one among
them.

\begin{table}[tbh]
\caption{\label{tab:Notation-and-symbols}Notations and symbols used throughout
this paper.}

~~~~~~~~~~~~~~~~~~~~~~%
\begin{tabular}{|c|>{\centering}p{10cm}|}
\hline
$N$ & Number of users\tabularnewline
\hline
$K$ & Number of REs (time-frequency slots)\tabularnewline
\hline
$M$ (or $M_{N}$,$M_{K}$) & Number of best (or worst) REs for each user. Denoted $M_{N}$ (or
$M_{K}$) where the dependence on $N$ (or $K$) is relevant\tabularnewline
\hline
$\mathbf{a}$ & Strategy profile (RE choices)\tabularnewline
\hline
$\mathbf{a}_{-n}$ & Strategy profile of user $n$'s rivals\tabularnewline
\hline
$a_{n}$ & User $n$'s action (RE)\tabularnewline
\hline
$I_{k}\left(\mathbf{a}_{-n}\right)$ & The interference user $n$ measures in RE $k$, given $\mathbf{a}_{-n}$\tabularnewline
\hline
$u_{n}$ & The utility function of user $n$\tabularnewline
\hline
$h_{n,m,k}$ & Channel gain between user $n$'s transmitter to user $m$'s receiver,
in RE $k$\tabularnewline
\hline
$P_{n}$ & User $n$'s transmission power\tabularnewline
\hline
$X_{\left(i\right)}$ & The $i$-th smallest variable among $X_{1},...,X_{N}$\tabularnewline
\hline
$\plim$ & Convergence in probability\tabularnewline
\hline
\end{tabular}
\end{table}

\subsection{Convergence to a PNE in a Frequency-Selective Interference Game}

Our games of interest are static games with static equilibria. However,
the process of converging to these static equilibria in practice is
of course dynamic. By designing a game with a PPoA close to one, we
are guaranteed that the performance of any distributed channel allocation
algorithm that converges to a PNE will be close to optimal. Therefore
the algorithm that we are looking for is not tailored to our specific
problem but rather has general properties of convergence to NE. A
great deal of work has been done on learning algorithms for NE (see
\cite{Babichenko2012,Pradelski2012,Shamma2005} and \cite{Rose2011}
for a summary). One of the best known candidates for this task is
the Fictitious Play (FP) algorithm \cite{Brown1951,Levin}. In FP,
each player keeps the empirical mean vectors of the frequencies each
other player has played his actions, and plays his best response to
this fictitious strategy.

Although it has a strong connection to NE, FP is not guaranteed to
converge at all. Convergence has only been proven for some special
games that do not include our game (for example see \cite{Perlaza2011}).
Even if FP converges, it may be to a mixed NE, which is undesirable
as a channel allocation solution. Proving the convergence of best-response
like dynamics (such as the FP) to a PNE in interference games is a
challenging task that is outside the scope of this paper. However,
our recent results \cite{Bistritz2016,Bistritz2018} show that convergence
of approximate best-response dynamics in interference games happens
with high probability, despite the fact that they are not potential
games.

A fundamental problem when implementing FP is the information it requires.
In a wireless network, not only does a user have hardly any information
about the previous action of each other user, but he also barely knows
how many users there are. This is the reason we require from our designed
utility of the FSIG (in \eqref{eq:2}) that the effect of the other
users on the utility of a user can be measured by measuring the interference.
Exploiting this property, in Algorithm \ref{alg:Modified-Fictitious-Play}
we adjust the FP to the wireless environment by modifying it such
that each user keeps track of a fictitious utility vector instead
of the empirical mean vector of the rivals' strategy profiles.

Additionally, we provide a simple mechanism to improve the chances
of convergence to a PNE. The strategy profile determines the interference,
but knowing the interference will not reveal the strategy profile.
Nevertheless, the continuity of the random channel gains suggests
that for each user, the interference vector is different for different
strategy profiles with probability 1. Hence users can detect that
two strategy profiles are different based on their measured interference.
If a PNE is reached it is played repeatedly, so we can exploit this
fact and let the users check for convergence to a PNE after enough
time, and set their fictitious utilities to zero if a PNE has not
been reached. This gives Part d of Algorithm \ref{alg:Modified-Fictitious-Play}.

We assume time is divided into discrete turns, and that in each turn
users choose actions simultaneously. The synchronization between the
users is only assumed to simplify the presentation. In fact, rarely
the dynamics of a game have to be synchronized in order to converge.
We demonstrate in simulations the behavior of our algorithm in an
asynchronized environment.

The Modified FP is described in detail in Algorithm \ref{alg:Modified-Fictitious-Play},
and its properties are summarized in the next proposition. Note that
$\alpha=1$ corresponds to the best-response dynamics.
\begin{prop}
Consider a FSIG $G$ where the $N$ users play according to Algorithm
\ref{alg:Modified-Fictitious-Play}. If $\alpha=\frac{1}{t+1}$ or
$\alpha$ is constant with time then
\begin{enumerate}
\item If a PNE is attained at $t_{0}$ it will be played for all $t>t_{0}$.
\item If the fictitious utility vectors converge, then the resulting strategy
profile is a NE.
\begin{enumerate}
\item If $\alpha=\frac{1}{t+1}$ then this might be a mixed NE.
\item If $\alpha$ is constant with time , then this is a PNE.
\end{enumerate}
\item If $\mathbf{a\in}A_{1}\mathbf{\times...\times}A_{N}$ is played for
every $t>t_{1}$ then $\mathbf{a}$ is a PNE.
\end{enumerate}
\end{prop}
\begin{IEEEproof}
Assume we are at turn $t=T$ and define $p_{i}=\sum_{t=1}^{T}\frac{I(\mathbf{a}_{-n}(t)=\mathbf{a}_{i,-n})}{T}$
for the rivals' strategy profile $\mathbf{a}_{i,-n}$, where $I$
is the indicator function. For $\alpha=\frac{1}{t+1}$, the equivalence
of the modified FP to the original FP immediately follows from
\begin{equation}
\sum_{i}p_{i}u_{n}(a_{n},\mathbf{a}_{i,-n})=\sum_{i}p_{i}u_{n}\left(a_{n},I\left(\mathbf{a}_{i,-n}\right)\right)=\frac{1}{T}\sum_{t=1}^{T}u_{n}\left(a_{n},I\left(\mathbf{a}_{-n}\left(t\right)\right)\right)\label{eq:33}
\end{equation}
since $\sum_{i}p_{i}u_{n}(a_{n},\mathbf{a}_{i,-n})$ is the mean empirical
utility according to the fictitious rival profile. Then, the result
follows from \cite{Levin}.

Consider next the case of a constant $\alpha$. If a PNE $\mathbf{a^{*}}$
is attained at $t_{0}$ then $a_{n}^{*}(t_{0})=\arg\underset{k}{\max}\,u_{n,k}(t_{0})$
and also $a_{n}^{*}(t_{0})=\arg\underset{k}{\max}\,\overline{U}_{n,k}(t_{0}-1)$
for each $n\in\mathcal{N}$. Considering the update rule and because
$a_{n}^{*}(t_{0})=\arg\underset{k}{\max\,}\overline{U}_{n,k}(t_{0}-1)=\arg\underset{k}{\max\,}u_{n,k}(t_{0})$
we get\footnote{For the proof it is enough to break ties in $\overline{U}_{n,k}(t)$
by choosing the previous action if it is maximal; otherwise break
ties arbitrarily. In Step d of the Modified FP break ties at random. }
\begin{equation}
\arg\underset{k}{\max}\,(1-\alpha)\overline{U}_{n,k}(t_{0}-1)+\arg\underset{k}{\max}\,\alpha u_{n,k}(t_{0})=\arg\underset{k}{\max\,}\overline{U}_{n,k}(t_{0})=a_{n}^{*}(t_{0}+1)\label{eq:34}
\end{equation}
and so on, for each $t>t_{0}.$ If the fictitious utility vectors
converge, then $\underset{t\rightarrow\infty}{\lim}\overline{U}_{n,k}(t)$
exists and is finite for each $k$ and $n$. From the update rule
we get $\alpha\underset{t\rightarrow\infty}{\lim}\overline{U}_{n,k}(t)=\alpha\underset{t\rightarrow\infty}{\lim}u_{n,k}(t)$
for each $n,k$ which means $\underset{t\rightarrow\infty}{\lim}\overline{U}_{n,k}(t)=\underset{t\rightarrow\infty}{\lim}u_{n,k}(t)$
for constant $\alpha$. Consequently, for all $t>t_{1}$ for some
large enough $t_{1}$, $a_{n}(t)=\arg\underset{k}{\max}\,\overline{U}_{n,k}(t)=\arg\underset{k}{\max}\,u_{n,k}(t)$
for each $n\in\mathcal{N}$ and hence $\mathbf{a}$ is a PNE.
\end{IEEEproof}
In Section \ref{sec:Simulation-Results} we show numerically that
the modified FP algorithm introduced in this section leads to a very
fast convergence to a PNE in the M-FSIG.

\subsection{Asymptotic NE Analysis in the Number of Users}

In this paper we analyze the probability for the existence of only
good NE asymptotically in the number of users. This should not be
confused with requiring $N$ (and $K$) to be extremely large. The
right way to interpret our approach is - given a finite $N$ and an
interference network of interest, what is the probability that our
designed game exhibits only good NE? the larger $N$ the larger this
probability is. Since a distribution over the ensemble of interference
networks is a formal way of counting networks, an equivalent interpretation
is - how many of the interference networks of size $N$ have only
good NE in our designed game? Throughout this work, all of the proofs
involve bounding the relevant probabilities for finite $N$. The values
of $N$ for which close to optimal performance is guaranteed are in
fact very reasonable. Already for $N=50$ the asymptotic effects take
place, as can be verified by our bounds for finite $N$ and also via
simulations (which suggest that even smaller values are enough).

Since we assumed that the number of users equals to the number of
REs (i.e, $N=K$), taking the number of users to infinity also takes
the number of REs to infinity. This does not mean that the bandwidth
of each user needs to decrease with $N$. We do not deal with networks
that are growing larger but with a fixed given large network, with
a fixed bandwidth assignment for each channel. For those networks
we bound from below the probability that only good NE exist. Although
our channel allocation does not assume anything about the associated
bandwidth considerations, we propose the following thought experiment
to demonstrate why large networks, that use more bandwidth in total,
always exist from a theoretical perspective. This is in addition to
their practical existence, especially in sensor networks (see \cite{Akyildiz2002}).

Assume that ten operators have ten separated networks, operating in
the same geographical area, each consists of $K=N=10$ users and
REs. By joining their pool of REs together, due to the selectivity
of the channel, users of operator A might find some REs of operator
B to be much more attractive than their current ones, and vice versa.
This gain is known as multi-user diversity. Our results both formalize
this intuition and provide a new strong argument why large networks
are to be preferred. This joining will allow them for a simple channel
allocation algorithm that requires no communication or overhead from
the network and achieves close to optimal performance. None of them
could do so well on his own. The total bandwidth of the converged
network is ten times larger than each of the original ones. The resulting
network has $N=K=100$, where our asymptotic results are strongly
valid.

Our approach can also be utilized to save bandwidth. Without a simple
distributed solution that allows for a close to optimal channel allocation,
a network operator might have to settle for an arbitrary allocation
that ignores the selectivity of the channel. The optimal solution
yields approximately twice the sum-rate obtained by a random allocation
(for reasonable $N$ values, as observed in our simulations). If operators
are interested in preserving their users demand, each of them can
use only half of its bandwidth and obtain the same performance in
the converged network.

This idea of converging networks demonstrates that large networks
(in terms of both $N$ and $K$) do not necessarily imply a smaller
bandwidth for each user. Of course that large networks do not really
have to be created from the convergence of smaller networks. This
is a theoretic idea that emphasizes the importance of viewing all
users and REs as though they all belong to a single large network.
It is analogous to the idea of analyzing the capacity of an interference
channel as though the whole bandwidth is available, where in practice
the spectrum of interest may be licensed to different operators.

\subsection{\label{Bipartite Graphs}The Bipartite Graph of Users and Resources}

The structure of our random NE is closely related to the structure
of perfect matchings in random bipartite graphs. In this subsection
we specify this bipartite graph and define the relevant terminology.

The key to observing the bipartite graph structure hidden in our game
is by simply looking at the $M$ best (or worst) REs of each user,
as determined by his direct channel gains alone. Throughout the paper,
we denote by $h_{n,n,(N-i+1)}$ the channel gain with the $i$-th
largest absolute value, which we refer to as the $i$-th best RE for
user $n$ (so $h_{n,n,(1)}$ is the worst RE).

This information is conveniently described using a bipartite graph,
defined as follows.
\begin{defn}
\label{def:Player-Channel Graph}Let $B_{M}=\left(\mathcal{N},\mathcal{K},E\right)$
be a user-resource bipartite graph, where $\mathcal{N}$ is the set
of user vertices, $\mathcal{K}$ is the set of resource vertices,
and $E$ is the set of edges between them. An edge $e$ between user
$n$ and RE $k$ exists ($e\in E$) if and only if RE $k$ is one
of the $M$ best (or worst) REs of user $n$, i.e., $\left|h_{n,n,k}\right|\geq\left|h_{n,n,\left(N-M+1\right)}\right|$.
\begin{itemize}
\item The degree of a user or a RE is defined as the degree of its corresponding
vertex in the graph.
\item A matching in this graph is a subset of edges such that no more than
a single edge is connected to each user vertex or RE vertex.
\item A maximum matching is the matching with the maximal cardinality.
\item A perfect matching is a maximum matching where all user and RE vertices
are connected to a single edge (have degree one). It is feasible only
with a balanced bipartite graph, where $\left|\mathcal{N}\right|=\mathcal{\left|K\right|}$
($N=K$).
\end{itemize}
\end{defn}
Note that $B_{M}$ represents the $M$-best (or worst) REs of each
user, and not which RE is allocated to which user.

Since the channel gains are random, they induce a random bipartite
graph. This raises the question of the probability for a perfect matching
in this random bipartite graph. Due to the independency of channel
gains between users, the edges connected to different user vertices
are independent. However, the edges connected to the same user vertex
are dependent.

\begin{figure}[H]
\noindent\begin{minipage}[t]{1\columnwidth}%
~~~~~~~~~~~~~~~~~~~~~~~~~~~~~~~~~~~~~~~~~~~~~~~~~~~~~~~~~~\includegraphics[clip,width=4cm,height=4cm]{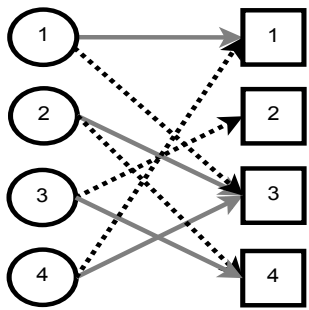}%
\end{minipage}\caption{\label{fig:bipartite}A toy example for a user-resource bipartite
graph with $N=K=4$.}
\end{figure}

Fig. \ref{fig:bipartite} presents a toy example for a user-resource
bipartite graph. Here each user vertex (a circle) is connected to
two RE vertices (squares). The dotted edges represent a possible perfect
matching in this graph. This perfect matching is a potential allocation
of REs to users where there is no resource sharing. Alternatively,
the gray edges represent another potential allocation (but not a matching)
where RE 3 is shared between User 2 and User 4.

\section{\label{Naive-FSIG section}The Naive Frequency-Selective Interference
Game Can Lead to Very Poor Performance}

Our performance metric is the (weighted) sum of achievable rates.
Hence, a natural choice for the utility of each user is his own term
in the sum - his achievable rate. In a strategic or economical environment,
this choice can be interpreted as the selfishness of the users. But
we deal with devices that implement a protocol, that like any other,
is wired in their circuits. This is a distributed protocol where each
device maximizes the utility function we chose as designers. Choosing
the achievable rate as the utility ignores the degree of freedom we
have in designing the protocol by designing the utility. Therefore
we name the resulting game the ``naive game''. We show in this section
that this naive choice for the utility can lead to poor global performance.
This motivates a more careful choice for the utility.
\begin{defn}
\textbf{\label{def:The Naive Game}The Naive Frequency-Selective Interference
Game (Naive-FSIG)} is a normal-form game with $N$ users as players,
where each has the set $A_{n}=\left\{ 1,2,...,K\right\} $ as a strategy
space. The utility function for user $n$ is
\begin{equation}
u_{n}\left(\mathbf{a}\right)=R_{n}\left(\mathbf{a}\right)=\log_{2}\left(1+\frac{P_{n}|h_{n,n,a_{n}}|^{2}}{N_{0}+I_{a_{n}}\left(\mathbf{a}_{-n}\right)}\right).\label{eq:3-1}
\end{equation}
\end{defn}
We show in this section that with strong enough interference, the
PPoA of the Naive-FSIG approach infinity (see Definition \ref{PPoA Definition}).
Strong interference is formally defined as follows
\begin{defn}
The strong interference condition on the channel gains requires that
for each $m$ and $l$
\begin{equation}
\frac{P_{m}}{N_{0}}\geq\underset{n,k}{\max}\frac{1}{\left|h_{m,n,l}\right|^{2}}\left(\frac{\left|h_{n,n,l}\right|^{2}}{\left|h_{n,n,k}\right|^{2}}-1\right).\label{eq:3}
\end{equation}
\end{defn}
Next we formalize the main theorem of this section. It shows that
the Naive-FSIG is a bad choice for a game formulation aimed to provide
a distributed solution for the channel allocation problem.
\begin{thm}[Naive-FSIG Main Theorem]
\label{NFSIG-PPOA} Let $\left\{ h_{n_{1},n_{2},k}\right\} $ form
an independent frequency-selective channel such that $\left|h_{n,n,1}\right|,\,...,\,\left|h_{n,n,k}\right|$
have a continuous distribution $F_{n}\left(x\right)$, with $F_{n}\left(x\right)>0$
for all $x>0$. If the strong interference condition \eqref{eq:3}
holds, then for all $\mu<1$ there are at least $\left(N^{\mu}\right)!$
PNE $\mathbf{a^{*}}$ such that
\begin{equation}
\plim_{N\to\infty}\frac{1}{N}\sum_{n}w_{n}R_{n}\left(\mathbf{a^{*}}\right)=0\label{eq:3-3}
\end{equation}
for any bounded weights $\left\{ w_{n}\right\} $. Specifically, the
PPoA of the Naive-FSIG approaches infinity in probability as $N\rightarrow\infty$.
\end{thm}
The proof of the above theorem follows by analyzing the PNE of the
Naive-FSIG for strong interference and evaluate the PPoA. The idea
of this analysis, which is the proof strategy of Theorem \ref{NFSIG-PPOA},
is explained here.

Trivially, a user who got his best RE without interference cannot
improve his utility. On the other hand, a user who is not in his best
RE (with the best channel gain) cannot necessarily improve his utility
if there are users in his more preferable REs. The influence of other
users in the RE of user $n$ on his utility is caused by the interference.
Consequently, the strength of the interference has a crucial effect
on the identity of the NE.

If the interference is strong enough, users in the same RE achieve
a negligible utility and the interference game becomes a ``collision
game''. In such a collision game, every permutation between users
and REs is a PNE, and every PNE is a permutation between users and
REs. We formalize this idea in Lemma \ref{NE are Permutations}.

The strong interference condition \eqref{eq:3} can also be written
as
\begin{equation}
\frac{P_{m}}{N_{0}}\left|h_{m,m,l}\right|^{2}\geq\underset{n,k}{\max}\frac{\left|h_{m,m,l}\right|^{2}}{\left|h_{m,n,l}\right|^{2}}\left(\frac{\left|h_{n,n,l}\right|^{2}}{\left|h_{n,n,k}\right|^{2}}-1\right)\label{eq:5-1}
\end{equation}
 for each $m,l$. For large $N$, we expect $\frac{\left|h_{n,n,l}\right|^{2}}{\left|h_{n,n,k}\right|^{2}}$
to be large due to the order statistics that stem from maximizing
over $n,k$. The rate of growth of the expectation due to order statistics
is bounded from above by $O\left(\sqrt{N}\right)$ (it is usually
much slower, e.g., $O\left(\ln N\right)$, see \cite{David1970}).
Hence, the strong interference condition is likely to hold when the
expectation of the ratio $\frac{\left|h_{m,m,l}\right|^{2}}{\left|h_{m,n,l}\right|^{2}}$
decreases with $N$. This is naturally possible if the area where
the users are located scales linearly with $N$ (constant user density).
In particular, $\frac{\left|h_{m,m,l}\right|^{2}}{\left|h_{m,n,l}\right|^{2}}$
can decrease linearly with $N$ if transmitter $m$ is close to receiver
$n$ and far from receiver $m$. In this case, a transmission power
that grows at least like $O\left(\sqrt{N}\right)$ is necessary to
obtain a reasonable SNR at the receiver, if the receiver is far away
from the transmitter. Alternatively, if $P_{m}$ is high enough then
\eqref{eq:5-1} also holds. Hence, we conclude that the strong enough
interference condition may hold due to geometrical considerations
or working in the high-SNR regime.

Now we need to analyze the performance of the worst perfect matching.
In Proposition \ref{Worst Channels are Asymptotically Worthless}
we show that a bad RE for a user can be asymptotically worthless,
i.e., results in an achievable rate that goes to zero. This Proposition
defines a bad RE as one of the worst $M_{N}$ such that $\underset{N\rightarrow\infty}{\lim}\frac{M_{N}}{N}=0$.
Next we answer the question of how many users can get such a bad RE
in a perfect matching. Unfortunately, all of them can. This is proved
in Theorem \ref{Existence of a Perfect Matching} which is based on
the theory of random bipartite graphs of users and REs introduced
in Subsection \ref{Bipartite Graphs}. This theory also allows us
to show that there are many such bad perfect matchings (Theorem \ref{Hall Jr}).
In summary, we show that for strong enough interference there are
many PNE (perfect matchings) which have a vanishing average achievable
rate per user. Thus, the PPoA of the Naive-FSIG approaches infinity
in the strong interference regime.

The following lemma shows that the set of PNE is the set of permutations
between users and REs.
\begin{lem}
\label{NE are Permutations} If the strong interference condition
\eqref{eq:3} holds, then the set of PNE of the Naive-FSIG is the
set of perfect matchings between users and REs, with cardinality $N!$.
\end{lem}
\begin{IEEEproof}
Rearranging the inequality condition we get $\frac{1}{N_{0}}\underset{k}{\min}\left|h_{n,n,k}\right|^{2}>\underset{l}{\max}\frac{\left|h_{n,n,l}\right|^{2}}{N_{0}+\underset{m}{\min}\left(\left|h_{m,n,l}\right|^{2}P_{m}\right)}$
for each $n$. This directly means that for every strategy profile
that is a permutation of users to REs, a user who deviates gets a
lower utility. The interpretation of these inequalities is that the
interference in the whole network is strong enough such that switching
to an occupied RE is always worse than staying with an interference-free
RE. Consequently, every permutation is an equilibrium. Conversely,
every equilibrium must be a permutation because all users prefer an
empty RE over a shared one (i.e. with positive interference).
\end{IEEEproof}
The lemma above implies that with strong enough interference, a PNE
of the Naive-FSIG may assign some users a bad RE. The next proposition
formulates this idea and shows that a bad RE can be asymptotically
worthless.
\begin{prop}
\label{Worst Channels are Asymptotically Worthless}Assume that $\left|h_{n,n,1}\right|,\,...,\,\left|h_{n,n,k}\right|$
are i.i.d. for each $n$, with a continuous distribution $F_{n}\left(x\right)$,
such that $F_{n}\left(x\right)>0$ for all $x>0$. Let $M_{N}$ be
a sequence such that $\underset{N\rightarrow\infty}{\lim}\frac{M_{N}}{N}=0$.
If $m\leq M_{N}$ then $\underset{n}{\max}\left|h_{n,n,(m)}\right|\rightarrow0$
in probability as $N\rightarrow\infty$.
\end{prop}
\begin{IEEEproof}
See Appendix \ref{Naive-FSIG Appendix}.
\end{IEEEproof}
Since $\frac{1}{N}\sum_{n}w_{n}R_{n}\left(\mathbf{a}\right)\leq w_{max}\underset{n}{\max\,}u_{n}$,
it follows from the lemma above that the users that are assigned one
of their $M_{N}$ worst channel gains have an average achievable rate
that converges to zero in probability. To evaluate the performance
of the worst PNE of the Naive-FSIG, we need to know how many users
can be assigned such a bad RE. Unfortunately, there is an $M_{N}$
such that $\underset{N\rightarrow\infty}{\lim}\frac{M_{N}}{N}=0$,
for which there exists a permutation between users and REs such that
each user gets one of his $M_{N}$ worst channel gains. Even worse,
there are actually many such permutations. This is shown by the following
theorem.
\begin{thm}
\label{Existence of a Perfect Matching}Let $B_{M_{N}}$ be the bipartite
user-resource graph of Definition \ref{def:Player-Channel Graph},
where users are connected to their $M_{N}$-worst REs. Let $\left\{ h_{n_{1},n_{2},k}\right\} $
form an independent frequency-selective channel.  If $M_{N}\geq(e+\varepsilon)\ln(N)$
for some $\varepsilon>0$, then the probability that a perfect matching
exists in $B_{M_{N}}$ approaches 1 as $N\rightarrow\infty$.
\end{thm}
\begin{IEEEproof}
See Appendix \ref{Existence Proof Appendix}.
\end{IEEEproof}
The condition $M_{N}\geq(e+\varepsilon)\ln(N)$ was necessary to ensure
that with high probability, a perfect matching of users to bad REs
exist. This large users' node degree has its own major effect on the
number of equilibria as well.
\begin{thm}[{Marshall Hall Jr \cite[Theorem 2]{HallJr1948}}]
 \label{Hall Jr}Suppose that $A_{1},A_{2},...,A_{N}$ are the finite
sets of desirable resources, i.e. user $n$ desires resource $a$
if and only if $a\in A_{n}$. If there exists a perfect matching between
users and resources and $\left|A_{n}\right|\geq M$ for $n=1\,,...,\,N$
where $M<N$, then the number of perfect matchings is at least $M!$.
\end{thm}
The main theorem of this section (Theorem \ref{NFSIG-PPOA}) readily
follows by joining together Theorem \ref{Hall Jr}, Lemma \ref{NE are Permutations},
Proposition \ref{Worst Channels are Asymptotically Worthless} and
Theorem \ref{Existence of a Perfect Matching} (choosing $M_{N}=N^{\mu}$
for some $\mu<1$).

\section{\label{M-FSIG Section}The M-Frequency Selective Interference Game
is Asymptotically Optimal}

The naive game introduced in the previous section has many bad equilibria
for strong enough interference. Hence, using its utility as a distributed
MAC protocol is a bad design choice. Our aim is to design an alternative
artificial utility function that obtains good global performance (see
Definition \ref{def:Global Performance}). The restriction on this
utility function is that it can be computed and maximized by each
user independently such that no communication between users is required.
This ensures that the resulting channel allocation protocol described
in Algorithm \ref{alg:Modified-Fictitious-Play} is fully distributed.

In general, it is intractable to find the optimal utility function
(in terms of Definition \ref{def:Global Performance}) without defining
a manageable domain of search (such as in pricing mechanisms). Fortunately,
in the channel allocation problem the optimal utility (asymptotically
in the number of users) happens to be a slight modification of the
naive selfish utility. In this new utility function defined below,
we choose the utility of each user to be greater than zero only for
his $M$ best REs, and be equal for them. We prove in this section
that this subtle change of the utility in Definition \ref{def:The Naive Game}
turns the tide for the performance of the PNE of the game. Instead
of many asymptotically worthless equilibria for strong enough interference,
we get that all equilibria are asymptotically optimal for any interference
regime. For this reason, in this game, the convergence to some PNE
is sufficient to provide asymptotically optimal global performance.
\begin{defn}
\textbf{The M-Frequency-Selective Interference Game (M-FSIG)} is a
normal-form game with parameter $M>0$ and $N$  users as players,
where each has the set $A_{n}=\left\{ 1,2,...,K\right\} $ as a strategy
space. The utility function for user $n$ is

\begin{equation}
u_{n}\left(\mathbf{a}\right)=\left\{ \begin{array}{cc}
\log_{2}\left(1+\frac{P_{n}|h_{n,n,(K-M+1)}|^{2}}{N_{0}+I_{a_{n}}\left(\mathbf{a}_{-n}\right)}\right) & \frac{\left|h_{n,n,a_{n}}\right|}{\left|h_{n,n,(K-M+1)}\right|}\geq1\\
0 & else
\end{array}\right..\label{eq:12}
\end{equation}
\end{defn}
Define the set of indices of the $M$ best channel gains of user $n$
by $\mathcal{M}_{n}=\left\{ k\,|\,\frac{\left|h_{n,n,k}\right|}{\left|h_{n,n,(K-M+1)}\right|}\geq1\right\} .$
Note that because $\frac{P_{n}|h_{n,n,(K-M+1)}|^{2}}{N_{0}+I_{a_{n}}\left(\mathbf{a}_{-n}\right)}>0$
for each $n\in\mathcal{N}$ with probability 1, user $n$ will never
choose a RE outside $\mathcal{M}_{n}$. This means that this utility
causes users to practically limit their chosen strategies to the smaller
set of their $M$-best REs.

Also note that due to the replacement of $h_{n,n,a_{n}}$ by $h_{n,n,(K-M+1)}$
in the utility, we obtain that
\begin{equation}
\arg\underset{a_{n}}{\max\,}u_{n}\left(\mathbf{a}\right)=\arg\underset{a_{n}\in\mathcal{M}_{n}}{\min}I_{a_{n}}\left(\mathbf{a}_{-n}\right).\label{eq:13}
\end{equation}
Hence in the M-FSIG each user $n\in\mathcal{N}$ only accesses REs
in $\mathcal{M}_{n}$ and prefers those with smaller interference.

We refer to this designed utility as a slight modification of the
utility of the Naive-FSIG (given in Definition \ref{def:The Naive Game})
since for each one of the $M$-best REs, the ratio of \eqref{eq:3-1}
and \eqref{eq:12} approaches one as $N$ approaches infinity. This
is due to $\frac{\left|h_{n,n,(K-M+1)}\right|}{\left|h_{n,n,(K)}\right|}\rightarrow1$
as $N\rightarrow\infty$ for a broad class of fading distributions,
as we prove in this section. Somewhat ironically, although in this
game each user is not exactly maximizing his own achievable rate,
his achievable right in equilibrium is going to be better maximized
than in the Naive-FSIG.

The above designed utility requires less information than a utility
that considers only REs that are better than some threshold (as in
\cite{Naparstek2013}). A threshold that avoids asymptotic performance
loss has to depend on $K$, where the dependence is determined by
the fading distribution. However, the fading distribution is not likely
to be known to the users. Furthermore, the above utility only requires
each user to track a small number of REs ($O\left(\ln K\right)$ instead
of $K$).

Naturally, the asymptotic optimality of the M-FSIG depends on the
distribution of the channel gains. In order to capture the general
nature of fading distributions, we define the following class. The
properties of these distributions that are useful for this section's
results are analyzed in Appendix \ref{sec:Exponentially-Dominated-Tail-Dis}.
\begin{defn}[Exponentially-Dominated Tail Distribution]
 \label{Exponentially-Dominated Tail}Let $X$ be a random variable
with a continuous CDF $F_{X}$. We say that $X$ has an exponentially-dominated
tail distribution if there exist $\alpha>0,\beta\in\mathbb{R\text{ }},\lambda>0,\gamma>0$
such that
\begin{equation}
\underset{x\rightarrow\infty}{\lim}\frac{1-F_{X}\left(x\right)}{\alpha x^{\beta}e^{-\lambda x^{\gamma}}}=1.\label{eq:17}
\end{equation}
\end{defn}
The main theorem of this section shows that the pure price of anarchy
of the M-FSIG is asymptotically optimal for fading coefficients with
an exponentially-dominated tail, under the right choice of the parameter
$M$, defined as follows.
\begin{thm}[M-FSIG Main Theorem]
 \label{M-FSIG Theorem} Let $\left\{ h_{n_{1},n_{2},k}\right\} $
form an $m$-dependent frequency-selective channel. If, for each $k$,
$\left|h_{n,n,k}\right|$ has an exponentially-dominated tail and
$M=\left(m+1\right)(e+\varepsilon)\ln K$ for some $\varepsilon>0$
then
\[
\ensuremath{\plim_{N\to\infty}}\frac{\underset{\mathbf{a}\in A_{1}\times...\times A_{N}}{\max}W(\mathbf{a})}{\underset{\mathbf{a}\in\mathcal{P}_{e}}{\min}W(\mathbf{a})}=1
\]
where $\mathcal{P}_{e}$ is the set of pure NE of the M-FSIG and $W(\mathbf{a})=\sum_{n}w_{n}R_{n}\left(\mathbf{a}\right)$
is the weighted sum of achievable rates with bounded weights (see
Definition \ref{def:Global Performance}).
\end{thm}
\begin{IEEEproof}
From Corollary \ref{Existence of an Asymptotically Optimal Permutation Equilibrium }
we know that for $M=\left(m+1\right)\left(e+\varepsilon\right)\ln K$,
there asymptotically exist many PNE in the M-FSIG, with high probability.
Denote by $\mathcal{S_{\tilde{\mathbf{a}}}}$ the set of sharing users
in $\tilde{\mathbf{a}}=\arg\underset{\mathbf{a}\in\mathcal{P}_{e}}{\min}W(\mathbf{a})$.
From Proposition \ref{why order statistics proposition} we know that
for every $\mathbf{a}^{*}\in\mathcal{P}_{e}$
\begin{multline}
\frac{W(\mathbf{\mathbf{a}^{*}})}{\underset{\mathbf{a}\in A_{1}\times...\times A_{N}}{\max}W(\mathbf{a})}\geq\frac{\sum_{n\in\mathcal{\mathcal{N}\setminus\mathcal{S_{\tilde{\mathbf{a}}}}}}w_{n}\log_{2}\left(1+\frac{P_{n}}{N_{0}}\left|h_{n,n,(K-M+1)}\right|^{2}\right)}{\sum_{n\in\mathcal{\mathcal{N}}}w_{n}\log_{2}\left(1+\frac{P_{n}}{N_{0}}\left|h_{n,n,(K)}\right|^{2}\right)}\geq\\
\frac{\sum_{n\in\mathcal{\mathcal{N}\setminus\mathcal{S_{\tilde{\mathbf{a}}}}}}w_{n}\underset{m}{\min}\log_{2}\left(1+\frac{P_{m}}{N_{0}}\left|h_{m,m,(K-M+1)}\right|^{2}\right)}{\sum_{n\in\mathcal{\mathcal{N}}}w_{n}\underset{m}{\max}\log_{2}\left(1+\frac{P_{m}}{N_{0}}\left|h_{m,m,(K)}\right|^{2}\right)}=\underbrace{\frac{\sum_{n\in\mathcal{\mathcal{N}\setminus\mathcal{S_{\tilde{\mathbf{a}}}}}}w_{n}}{\sum_{n\in\mathcal{\mathcal{N}}}w_{n}}}_{A}\underbrace{\frac{\underset{m}{\min}\log_{2}\left(1+\frac{P_{m}}{N_{0}}\left|h_{m,m,(K-M+1)}\right|^{2}\right)}{\underset{m}{\max}\log_{2}\left(1+\frac{P_{m}}{N_{0}}\left|h_{m,m,(K)}\right|^{2}\right)}}_{B}\underset{\begin{array}{c}
N\rightarrow\infty\\
(a)
\end{array}}{\rightarrow}1\label{eq:30}
\end{multline}
where (a) follows since for $M=\left(m+1\right)\left(e+\varepsilon\right)\ln K$:
\begin{itemize}
\item In Theorem \ref{Non-Existence of Bad Equilibria } (Section V.I) we
prove that for every PNE of the M-FSIG , and specifically the worst
one, $\frac{\left|\mathcal{\mathcal{S_{\tilde{\mathbf{a}}}}}\right|}{N}\rightarrow0$
as $N\rightarrow\infty$. Hence $A\rightarrow1$ as $N\rightarrow\infty$,
due to
\begin{equation}
1\geq\frac{\sum_{n\in\mathcal{\mathcal{N}\setminus\mathcal{S_{\tilde{\mathbf{a}}}}}}w_{n}}{\sum_{n\in\mathcal{\mathcal{N}}}w_{n}}=A=1-\frac{\sum_{n\in\mathcal{\mathcal{S_{\tilde{\mathbf{a}}}}}}w_{n}}{\sum_{n\in\mathcal{\mathcal{N}}}w_{n}}\geq1-\frac{\left|\mathcal{\mathcal{S_{\tilde{\mathbf{a}}}}}\right|}{N}\frac{w_{\max}}{w_{\min}}\underset{\begin{array}{c}
N\rightarrow\infty\end{array}}{\rightarrow}1.\label{eq:31}
\end{equation}
\item In Theorem \ref{Max-Min Fairness} (Section V.II) we prove that $B\rightarrow1$
as $N\rightarrow\infty$.
\end{itemize}
\end{IEEEproof}
In the following two subsections we prove Theorem \ref{Non-Existence of Bad Equilibria }
and Theorem \ref{Max-Min Fairness} that are necessary for our main
Theorem above. The first subsection is about the structure of the
PNE of the M-FSIG. We show in Theorem \ref{Non-Existence of Bad Equilibria }
that all of the PNE of the M-FSIG are almost a perfect matching in
between users and $M$-best REs, so almost all users obtain an interference-free
RE. The second subsection is about the weighted sum-rate performance
of the PNE of the M-FSIG. We show in Theorem \ref{Max-Min Fairness}
that for exponentially-dominated tail distributions and $m$-dependent
REs, in all PNE of the M-FSIG, even the minimal rate of a user is
asymptotically optimal, provided that this user is in an interference-free
RE.

\subsection{Asymptotic Existence and Structure of the Equilibria}

In this subsection we argue about the structure of the PNE of the
M-FSIG using two complementary results. The first shows that there
are many PNE which are perfect matchings between users and $M$-best
REs. The second shows that a PNE that is far from a perfect matching
cannot asymptotically exist. Together we conclude that all PNE are
almost \textquotedbl at least\textquotedbl{} a perfect matching.

Using the existence Theorem for a perfect matching in Appendix \ref{Existence Proof Appendix}
together with Theorem \ref{Hall Jr} we get the following corollary.
\begin{cor}
\label{Existence of an Asymptotically Optimal Permutation Equilibrium }Let
$B_{M}$ be the bipartite user-resource graph of Definition \ref{def:Player-Channel Graph},
where users are connected to their $M$-best REs. Let $\left\{ h_{n_{1},n_{2},k}\right\} $
form an $m$-dependent frequency-selective channel. If the M-FSIG
parameter is chosen such that $M\geq\left(e+\varepsilon\right)\left(m+1\right)\ln(N)$
for some $\varepsilon>0$, then the probability that there are at
least $M!$ perfect matchings in $B_{M}$ approaches 1 as $N\rightarrow\infty$.
\end{cor}
It turns out that the asymptotically optimal PNE are typical equilibria
for this game; in other words all other equilibria have almost the
same asymptotic structure, which is a perfect matching. This property
eases the requirements for the dynamics and allows simpler convergence
with good performance.
\begin{defn}
Define a shared RE as a RE that is chosen by more than one user. Define
a sharing user as a user that chose a shared RE.
\end{defn}

\begin{thm}
\label{Non-Existence of Bad Equilibria }Let $\left\{ h_{n_{1},n_{2},k}\right\} $
form an $m$-dependent frequency-selective channel. Suppose that $M\geq\left(e+\varepsilon\right)\left(m+1\right)\ln(N)$
for some $\varepsilon>0$. If \textbf{$\mathbf{a^{*}}$} is a PNE
of the M-FSIG with $N_{c}$ sharing users, then $\frac{N_{c}}{N}\rightarrow0$
in probability as $N\rightarrow\infty$. Furthermore\footnote{This statement is stronger since it argues about the probability that
all PNE will have $\frac{N_{c}\left(\mathbf{a^{*}}\right)}{N}\rightarrow0$
as $N\rightarrow\infty$.}, $\underset{\mathbf{a^{*}}\in\mathcal{P}_{e}}{\max}\frac{N_{c}\left(\mathbf{a^{*}}\right)}{N}\rightarrow0$
in probability as $N\rightarrow\infty$, where $\mathcal{P}_{e}$
is the set of pure NE.
\end{thm}
\begin{IEEEproof}
See Appendix \ref{M-FSIG Appendix}.
\end{IEEEproof}
By the definition of the M-FSIG, the achievable rate of a non-sharing
user is close to optimal (attained in the best REs for zero interference).
According to Theorem \ref{Non-Existence of Bad Equilibria } above,
almost all of the users are non-sharing users. Consequently, almost
all users have almost optimal performance. On the other hand, the
sharing users do not necessarily suffer from poor performance: their
chosen RE, although shared, has the minimal interference among the
$M$ available REs. If $M$ is increasing with $N$, this minimal
interference might not be large at all.

\subsection{Asymptotic Optimality of the Equilibria - Order Statistics of Fading
Channels}

In the last subsection we proved that all the PNE of the M-FSIG have
the same structure asymptotically, which is ``almost a perfect matching''
between users and REs, such that each user gets one of his $M$-best
REs. In this subsection we want to evaluate the global performance
(see Definition \ref{def:Global Performance}) of such an ``almost
perfect matching''. Since we measure the merits of the equilibria
via the pure price of anarchy, we are interested in where does $\frac{\underset{\mathbf{a}\in\mathcal{P}_{e}}{\min}W(\mathbf{a})}{\underset{\mathbf{a}\in A_{1}\times...\times A_{N}}{\max}W(\mathbf{a})}\rightarrow1$
as $N\rightarrow\infty$ for bounded weights (see Definition \ref{def:Bounded Weights}).

At first glance it may seem that such analysis would involve the identity
of the PNE and the optimal value of the global performance function
$W(\mathbf{a})$, and thus could be quite tiresome. Fortunately, the
definition of the M-FSIG and the ``almost perfect matching'' structure
of its PNE allow for a simpler approach.
\begin{prop}
\label{why order statistics proposition}Denote by $\mathcal{S_{\tilde{\mathbf{a}}}}$
the set of sharing users in $\tilde{\mathbf{a}}=\arg\underset{\mathbf{a}\in\mathcal{P}_{e}}{\min}W(\mathbf{a})$.
The PPoA of the M-FSIG satisfies
\begin{equation}
\frac{1}{PPoA}=\frac{\underset{\mathbf{a}\in\mathcal{P}_{e}}{\min}W(\mathbf{a})}{\underset{\mathbf{a}\in A_{1}\times...\times A_{N}}{\max}W(\mathbf{a})}\geq\frac{\sum_{n\in\mathcal{\mathcal{N}\setminus\mathcal{S_{\tilde{\mathbf{a}}}}}}w_{n}\log_{2}\left(1+\frac{P_{n}}{N_{0}}\left|h_{n,n,(K-M+1)}\right|^{2}\right)}{\sum_{n\in\mathcal{\mathcal{N}}}w_{n}\log_{2}\left(1+\frac{P_{n}}{N_{0}}\left|h_{n,n,(K)}\right|^{2}\right)}.\label{eq:16}
\end{equation}
\end{prop}
\begin{IEEEproof}
By definition of the M-FSIG, for each PNE $\mathbf{a^{*}}$ the inequality
$W(\mathbf{a^{*}})\geq\sum_{n\in\mathcal{\mathcal{N}\setminus S_{\mathbf{a^{*}}}}}w_{n}\log_{2}\left(1+\frac{P_{n}}{N_{0}}\left|h_{n,n,(K-M+1)}\right|^{2}\right)$
holds (note that $\mathcal{\mathcal{N}\setminus S_{\mathbf{a^{*}}}}$
is the set of non-sharing users in $\mathbf{a^{*}}$). By the definition
of the channel allocation problem $W(\mathbf{a})\leq\sum_{n\in\mathcal{N}}w_{n}\log_{2}\left(1+\frac{P_{n}}{N_{0}}\left|h_{n,n,(K)}\right|^{2}\right)$
holds for each $\mathbf{a}\in A_{1}\times...\times A_{N}$. By choosing
$\tilde{\mathbf{a}}=\arg\underset{\mathbf{a}\in\mathcal{P}_{e}}{\min}W(\mathbf{a})$
in the first inequality and $\mathbf{a_{opt}}=\arg\underset{\mathbf{a}\in A_{1}\times...\times A_{N}}{\max}W(\mathbf{a})$
in the second we reach our conclusion.
\end{IEEEproof}
The proposition above suggests that if $\left|h_{n,n,(K-M+1)}\right|$
is asymptotically close, in some sense, to $\left|h_{n,n,(K)}\right|$,
and $\sum_{n\in\mathcal{S_{\tilde{\mathbf{a}}}}}\left|h_{n,n,(K)}\right|$
is asymptotically negligible compared to $\sum_{n\in\mathcal{N}\setminus\mathcal{S_{\tilde{\mathbf{a}}}}}\left|h_{n,n,(K)}\right|$,
then the PPoA of the M-FSIG converges to 1 in probability as $K\rightarrow\infty$.
This leads us to explore the asymptotic statistical behavior of $|h_{n,n,(K-M+1)}|$
and $|h_{n,n,(K)}|$. Therefore a statistical equilibrium analysis
is replaced by an order statistics analysis of the channel gains,
which is much simpler.

For each $n\in\mathcal{N}$, Let $X_{n,1},X_{n,2},...,X_{n,K}$ be
a sequence of random variables. We denote by $X_{n,(i)}$ the $i$-th
variable in the sorted list of $X_{n,1},X_{n,2},...,X_{n,K}$, with
increasing order, i.e. $X_{n,(1)}\leq X{}_{n,(2)}\leq...\leq X{}_{n,(K)}$.
The statistics of $X_{n,(i)}$ are called the order statistics. We
are interested in the statistics of $X_{n,(K-M_{K}+1)}$ for a sequence
$M_{K}$ such that $\underset{K\rightarrow\infty}{\lim}\frac{M_{K}}{K}=0$
and $\underset{K\rightarrow\infty}{\lim}M_{K}=\infty$. (i.e. intermediate
statistics) versus those of $X_{n,(K)}$ (i.e. extreme statistics).
We assume for simplicity that $\left\{ X_{n,k}\right\} $ are identically
distributed; hence, we choose them as normalized channel gains.

Now we turn to characterize the distributions that we are interested
in. The first distinction we have to make is between bounded and unbounded
random variables.
\begin{rem}[Bounded Random Variables]
 If $F_{X}^{-1}(1)<\infty$ then $X$ is a bounded random variable.
In that case the behavior of $X_{n,(K-M_{K}+1)}$ compared to that
of $X_{n,(K)}$ for a sequence $M_{K}$ such that $\underset{K\rightarrow\infty}{\lim}\frac{M_{K}}{K}=0$
and $\underset{K\rightarrow\infty}{\lim}M_{K}=\infty$ is much simpler
to analyze. By substituting $\left|h_{n,n,k}\right|=F_{X}^{-1}(1)-X_{n,k}$
in Proposition \ref{Worst Channels are Asymptotically Worthless},
(assuming independent variables and a continuous distribution) we
obtain $\underset{n}{\max}\left(F_{X}^{-1}(1)-X_{n,(K-M_{K}+1)}\right)\rightarrow0$
as $K\rightarrow\infty$, hence $X_{n,(K-M_{K}+1)}-X_{n,(K)}\rightarrow0$
for all $n$ in probability as $K\rightarrow\infty$. This immediately
validates all the results of this section for bounded random variables.
\end{rem}
Keeping the above remark in mind, we formulate our results assuming
unbounded variables. Note that all the classical fading distributions
are indeed unbounded. We are interested in proving our results for
a large class of fading distributions, which evidently tend to belong
to the family of exponentially-dominated tail distributions (see Definition
\ref{Exponentially-Dominated Tail}). Appendix \ref{sec:Exponentially-Dominated-Tail-Dis}
provides the necessary properties of exponentially-dominated distributions
that we use for our following results. Note that the following results
are true for $m$-dependent REs, that includes i.i.d. REs as a special
case.

We start by proving that, for exponentially-dominated tail distributions,
the ratio of the intermediate statistics to the extreme statistics
approaches one as the number of REs approaches infinity. It shows
that for a slow enough increasing $M_{K}$, the interference-free
$M_{K}$ best RE of a user is asymptotically optimal in ratio.
\begin{thm}
\label{non-log Asymptotic Optimality}Let $X_{1},X_{2},...,X_{K}$
be unbounded $m$-dependent random variables with an exponentially-dominated
tail. Let $M_{K}$ be a sequence such that $\underset{K\rightarrow\infty}{\lim}M_{K}=\infty$.
If $\underset{K\rightarrow\infty}{\lim}\frac{M_{K}}{\ln^{\mu}K}=0$
for some $\mu<1$ then $\frac{X_{(K-M_{K}+1)}}{X_{(K)}}\rightarrow1$
in probability as $K\rightarrow\infty$.
\end{thm}
\begin{IEEEproof}
See Appendix \ref{M-FSIG Appendix}.
\end{IEEEproof}
The following simple proposition shows that under the conditions of
Theorem \ref{non-log Asymptotic Optimality}, the achievable rate
of the interference-free $M_{K}$ best RE is also asymptotically optimal
in ratio.
\begin{prop}
If $\frac{X_{(K-M_{K}+1)}}{X_{(K)}}\rightarrow1$ in probability as
$K\rightarrow\infty$ then for each $a>0$, $\frac{\log_{2}\left(1+\frac{P}{N_{0}}aX_{(K-M_{K}+1)}\right)}{\log_{2}\left(1+\frac{P}{N_{0}}aX_{(K)}\right)}\rightarrow1$
in probability as $K\rightarrow\infty$.
\end{prop}
\begin{IEEEproof}
From the monotonicity of $\frac{\log_{2}\left(1+x\right)}{x}$ for
$x>0$ it follows that $\frac{X_{(K-M_{K}+1)}}{X_{(K)}}\leq\frac{\log_{2}\left(1+\frac{P}{N_{0}}aX_{(K-M_{K}+1)}\right)}{\log_{2}\left(1+\frac{P}{N_{0}}aX_{(K)}\right)}$.
\end{IEEEproof}
Theorem \ref{non-log Asymptotic Optimality} implies that for almost
all users $\frac{X_{n,(K-M_{K}+1)}}{X_{n,(K)}}\rightarrow1$ as $K\rightarrow\infty$.
In fact, because our global performance measure is logarithmic with
respect to the channel gain power, a stronger result holds. The next
theorem suggests that every non-sharing user enjoys asymptotically
similar conditions. More formally, the minimal rate achieved by the
non-sharing users is asymptotically optimal (maximized), which is
known also as Max-min fairness between these non-sharing users. This
shows that the perfect matching equilibria of the M-FSIG has Max-min
fairness between all users. Note that the fact there is a perfect
matching with performance that approaches the optimal solution for
$W(\mathbf{a})$ means that a perfect matching can achieve the maximal
multi-user diversity.
\begin{thm}[Max-Min Fairness for non-sharing users]
\label{Max-Min Fairness} Let $\left\{ X_{n,k}\right\} $ be identically
distributed unbounded random variables with an exponentially-dominated
tail. Assume that $\left\{ X_{n,k}\right\} _{k}$ are $m$-dependent
for each $n$. Let $M_{K}$ be a sequence such that $\underset{K\rightarrow\infty}{\lim}M_{K}=\infty$.
Let $a_{n}$ be a positive sequence. If $\underset{K\rightarrow\infty}{\lim}\frac{M_{K}}{K^{\mu}}=0$
for some $\mu<1$ then
\[
\plim_{K\to\infty}\frac{\underset{n}{\min}\log_{2}\left(1+\frac{P_{n}}{N_{0}}a_{n}X_{n,(K-M_{K}+1)}\right)}{\underset{n}{\max}\log_{2}\left(1+\frac{P_{n}}{N_{0}}a_{n}X_{n,(K)}\right)}=1.
\]
\end{thm}
\begin{IEEEproof}
See Appendix \ref{M-FSIG Appendix}.
\end{IEEEproof}
The role of the sequence $a_{n}$ is to allow for $a_{n}X_{n,k}$
to have non identical parameters for different users, by choosing
$X_{n,k}=\frac{\left|h_{n,n,k}\right|^{2}}{a_{n}}$.

\section{Simulation Results\label{sec:Simulation-Results}}

Our NE analysis in this paper is probabilistic and asymptotic with
the number of users $N$. It provides bounds in terms of $N$ on the
probability that all NE are asymptotically optimal. Simulations provide
another assessment about of which $N$ values are enough for the asymptotic
effects to hold. Since some of our bounds are not tight, simulations
suggest that even smaller values of $N$ are enough.

In our simulations we used a Rayleigh fading network; i.e. $\left\{ \left|h_{m,n,k}\right|\right\} $
are independent Rayleigh random variables. Hence $\left\{ \left|h_{m,n,k}\right|^{2}\right\} $
are independent exponential random variables. The parameters $\left\{ \lambda_{m,n}\right\} $
for the exponential variables were chosen according to the users'
random positions such that $\lambda_{m,n}=\frac{G}{r_{m.n}^{\gamma}}$,
where $r_{m,n}$ is the distance between transmitter $m$ and receiver
$n$, $\gamma$ is the path-loss exponent (which is chosen to be $\gamma=3$),
and $G=\left(\frac{\lambda_{wave}}{4\pi}\right)^{\gamma}$ where $\lambda_{wave}$
is the wavelength (chosen to be $\lambda_{wave}=\frac{3\cdot10^{8}}{2.4\cdot10^{9}}$).
The users' receiver positions are chosen uniformly at random on a
disk with radius $R=1000[m]$, and each transmitter position is at
distance $R_{link}\sim N\left(50,25\right)$ (normally distributed)
and angle $\theta\sim U\left([0,2\pi]\right)$ from its intended receiver.
The transmission powers were chosen such that the mean SNR for each
link, in the absence of interference, is 15{[}dB{]}. Users play according
to the Modified FP algorithm (Algorithm \ref{alg:Modified-Fictitious-Play})
including Step d with $\tau=60$ and $\alpha=0.5$. The parameters
for the simulations are summarized in Table \ref{tab:Simulation-Network-Parameters}.

\begin{figure}[H]
\noindent\begin{minipage}[t]{1\columnwidth}%
~~~~~~~~~~~~~~~~~~~~~~~~~~~~~~~~~~~~~\includegraphics[clip,width=9cm,height=6cm]{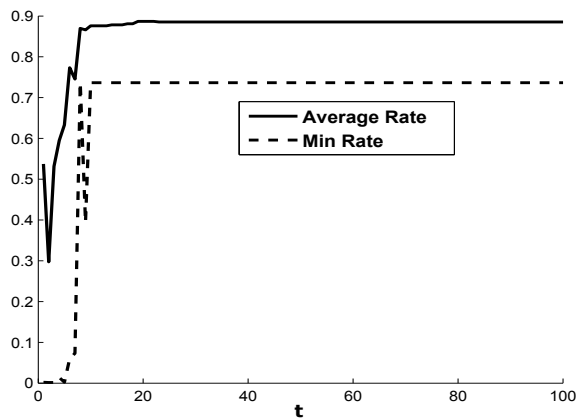}%
\end{minipage}\caption{\label{fig:Single}Sum-rate and min-rate compared to the optimal permutation
allocation for a single realization in the uncorrelated case.}
\end{figure}

In Fig. \ref{fig:Single} we present a typical convergence of the
Modified FP for a single game realization with $N=50$ and $M=8$.
Clearly convergence is very fast and occurs within 20 iterations.
The ratio of the sum of achievable rates to that of an optimal allocation
is close to 1, and the ratio of the minimal achievable rate is not
far behind. This corresponds to a convergence to a PNE with two sharing
users. These users do not decrease the minimal rate significantly,
as they are the best choice among $M=8$ for each other. The geometry
of the scenario allows for the minimal interferer to be far away.

\begin{figure}[H]
\noindent\begin{minipage}[t]{1\columnwidth}%
~~~~~~~~~~~~~~~~~~~~~~~~~~~~~~~~~~~~~\includegraphics[clip,width=9cm,height=6cm]{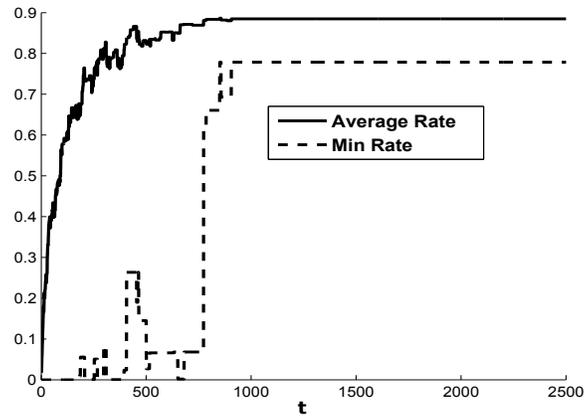}%
\end{minipage}\caption{\label{fig:SingleASync}Asynchronized modified FP convergence in a
single realization.}
\end{figure}

In Fig. \ref{fig:SingleASync} we present again a typical convergence
for a single game realization, but this time where users act asynchronously.
In each turn a user is chosen at random among all the $N$ users to
perform the Modified FP. In time $t$, the non-acting users keep their
variables, described in Algorithm \ref{alg:Modified-Fictitious-Play},
fixed. The algorithm converges to a PNE with a similar performance
and convergence rate to the synchronous case. Convergence time is
approximately 50 times larger simply because now only a single user
acts each turn instead of $N=50$ users. This figure demonstrates
that synchronization is not an issue for our channel allocation algorithm.

\begin{figure}[H]
\noindent\begin{minipage}[t]{1\columnwidth}%
~~~~~~~~~~~~~~~~~~~~~~~~~~~~~~~~~~~~~\includegraphics[clip,width=9cm,height=6cm]{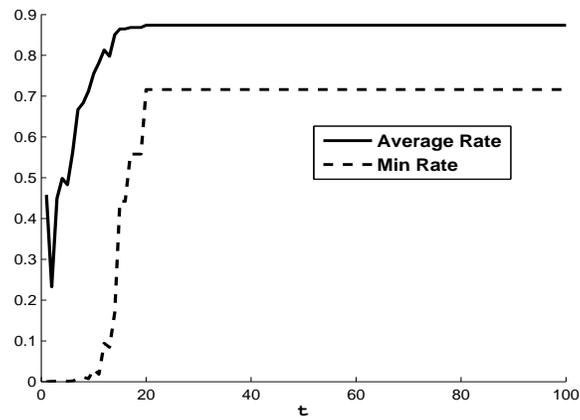}%
\end{minipage}\caption{\label{fig:SinglemDependent}Modified FP convergence in a single realization
in the $m$-dependent case.}
\end{figure}

Fig. \ref{fig:SinglemDependent} shows the convergence in a single
network realization, for $N=50$ and $M=12$, for the case of the
$m$-dependent frequency-selective channel. The direct channel gains
of each user were generated using the Extended Pedestrian A model
(EPA, see \cite{LTE2009}) for the excess tap delay and the relative
power of each tap. The parameter $m$ of their dependency is roughly
given by $m\geq\frac{T_{s}N}{50\sigma_{\tau}}$ where $T_{s}$ is
the duration of a symbol and $\sigma_{\tau}$ is the delay spread
of the channel, which is $143[nsec]$. We chose $T_{s}=0.44[\mu sec]$
so $m=3$. As expected, the behavior is very similar to the uncorrelated
case. The convergence time was unaffected and the sum and min rate
were only slightly reduced. Note that the resulting pure NE is a perfect
matching, (i.e., has no sharing users).

\begin{figure}[H]
\noindent\begin{minipage}[t]{1\columnwidth}%
~~~~~~~~~~~~~~~~~~~~~~~~~~~~~~~~~~~~~\includegraphics[clip,width=9cm,height=6cm]{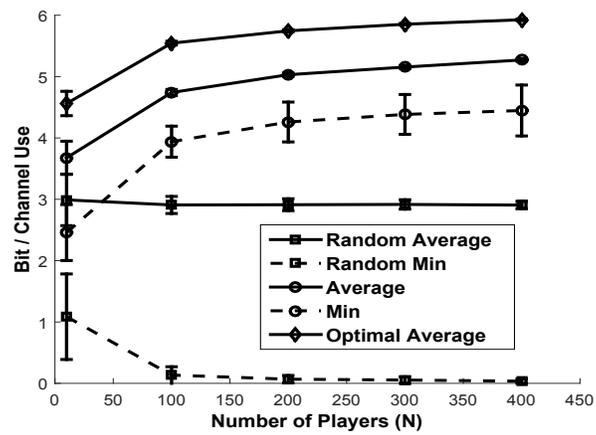}%
\end{minipage}

\caption{\label{fig:VersusN}Rates as a function of $N$ averaged over 100
realizations.}
\end{figure}

In Fig. \ref{fig:VersusN} we show the effect of the number of users
on the rates, with $M=\left\lceil 3\ln N\right\rceil $, averaged
over 100 realizations. Only for $N=10$ we chose $M=5$. We present
the average and minimal achievable rates, compared to the sum-rate
optimal permutation allocation and random permutation allocation mean
achievable rates. The benefit over a random permutation is significant,
especially in terms of the minimal rate. The rate increase is due
to the growing expected value of the best channel gain for each user.
This phenomenon (multi-user diversity) of course does not take place
for a random permutation. In a random permutation the average user
gets his median channel gain, and the minimal allocated channel gain
has a decreasing expectation. The random permutation may be interpreted
as a result of a channel allocation scheme that ignores the selectivity
of the channel, or as a random PNE of the Naive-FSIG for strong enough
interference. The standard deviations of the mean rates are small
as expected from the similarity of all NE, and the standard deviations
of the minimal rate are higher due to the changing number of sharing
users between different realizations.

\begin{figure}[H]
\noindent\begin{minipage}[t]{1\columnwidth}%
~~~~~~~~~~~~~~~~~~~~~~~~~~~~~~~~~~~~~\includegraphics[clip,width=9cm,height=6cm]{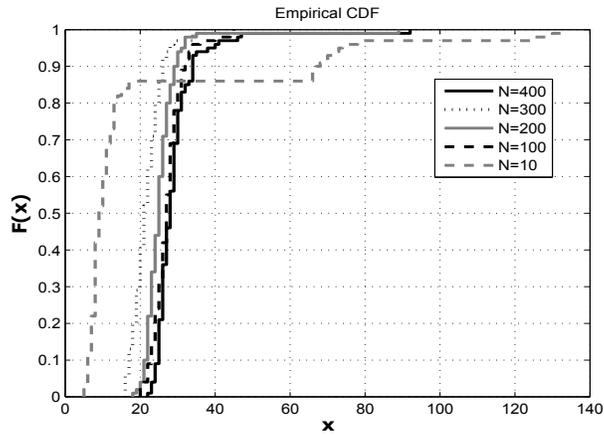}%
\end{minipage}

\caption{\label{fig:Convergence}Empirical cumulative distribution functions
of the convergence time for different $N$ values, over 100 realizations.}
\end{figure}

In Fig. \ref{fig:Convergence} we present the empirical cumulative
distribution functions of the modified FP convergence time, that were
derived from 100 realizations. Five different functions are depicted,
for $N=10,100,200,300,400$, and $K=N$. We see that about 90\% of
the dynamics converge in less than 40 iterations. We can also see
that the convergence time is very weakly affected by the number of
users, which means the modified FP has excellent scalability properties.

Interestingly, the two last figures suggest that the asymptotic effects
are starting to become valid already for small values of $N$. For
$N=10$, \textasciitilde 85\% of the realizations lead to convergence
within 20 iterations, which is typical for larger $N$ values. The
rest \textasciitilde 15\% of realizations lead to convergence within
140 iterations. This ``anomaly'' shrinks significantly for larger
$N$ values. The sum-rate was \textasciitilde 80\% of the optimal,
where multi-user diversity is still not significant for $N=10$. All
the resulting PNE consist of zero or two sharing users, which coincides
with the fact that all PNE are almost perfect matchings in the associated
bipartite graph.

\begin{figure}[H]
~~~~~~~~~~~~~~~~~~~~~~~~~~~~~~~~~~~~~\includegraphics[clip,width=9cm,height=6cm]{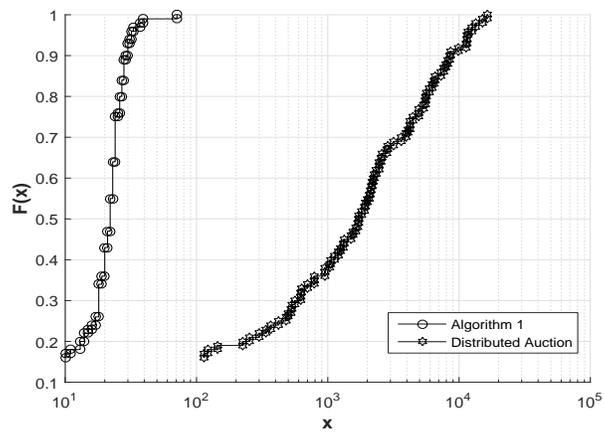}

\caption{\label{fig:ConvergenceComparison}Empirical cumulative distribution
functions of the convergence time for $N=125$ over 100 realizations,
compared to state of the art.}
\end{figure}

\begin{figure}[H]
\noindent\begin{minipage}[t]{1\columnwidth}%
~~~~~~~~~~~~~~~~~~~~~~~~~~~~~~~~~~~~~\includegraphics[clip,width=9cm,height=6cm]{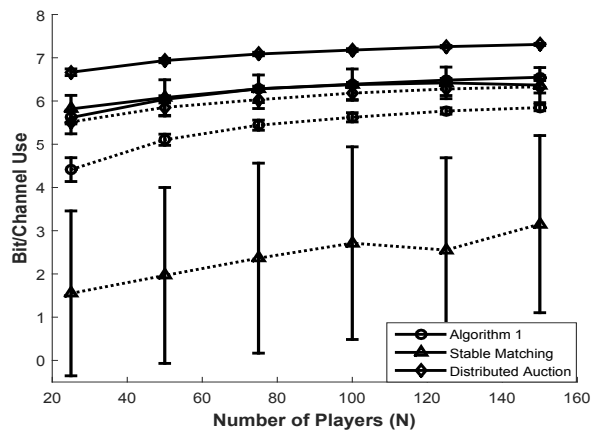}%
\end{minipage}

\caption{\label{fig:VersusNComparison}Rates as a function of $N$ averaged
over 100 realizations, compared to state of the art algorithms.}
\end{figure}

Finally, we compare the performance of Algorithm \ref{alg:Modified-Fictitious-Play}
to the distributed auction algorithm of \cite{Naparstek2014} and
the stable matching algorithm of \cite{Leshem2012}. In Fig. \ref{fig:VersusNComparison}
we present the mean and minimal rates of these algorithms as a function
of $N$, averaged over 100 realizations and for $N=25,50,75,100,125,150$.
In Fig. \ref{fig:ConvergenceComparison} we present the empirical
CDF of the convergence times for the experiment with $N=125$. The
distributed auction algorithm is $\varepsilon$-optimal and we chose
$\varepsilon=\frac{1}{N}$, so it naturally achieves the same rates
(sum-rate and minimal) as the Hungarian algorithm in previous simulations.
However, its convergence rate is more than one hundred times slower
than that of Algorithm \ref{alg:Modified-Fictitious-Play}. Since
a distributed channel allocation algorithm has to coverage within
the coherence time of the channel, this slow convergence time is very
restricting in practice. The stable matching algorithm always converges
within a single contention window, assuming synchronized users. Our
convergence time is naturally slower, but still very fast and does
not require synchronized users (as demonstrated in Fig. \ref{fig:SingleASync}).
Our sum-rate is very similar to that of the stable matching algorithm,
but the minimal rate is more than three times better. The stable matching
algorithm has no fairness guarantees, and as indicated by the large
error bars, results in a very different minimal rate in different
networks. In fact, it may even choose not to assign any REs to some
users, making the minimal rate zero. Nevertheless, these appealing
performance improvements are not the main advantage of our algorithm
over the existing ones. Both the distributed auction algorithm and
the stable matching require that all users can hear all other users
(fully connected network) so they can listen before transmission in
a CSMA manner. This significantly limits the distributed nature of
the network. In Algorithm \ref{alg:Modified-Fictitious-Play}, every
user transmits in the RE that suits his utility best without considering
any other user or avoiding a collision. Still, the overall performance
is not only not worse but significantly better. This strengthens our
claim that using utility design tools for distributed optimization
can lead to much better distributed algorithms.

\begin{table}[tbh]
\caption{\label{tab:Simulation-Network-Parameters}Simulation network parameters.}

~~~~~~~~~~~~~~~~~~~~~~~~~~%
\begin{tabular}{|c|>{\centering}p{8cm}|c|}
\hline
$\gamma$ & Path-loss exponent & 3\tabularnewline
\hline
$\lambda_{wave}$ & Wavelength (in meters) & $\frac{1}{2.4}10^{-9}$\tabularnewline
\hline
SNR & (in the absence of interference) & 15{[}dB{]}\tabularnewline
\hline
$R$ & Radius of the disk region (in meters) & $1000$\tabularnewline
\hline
$R_{link}$ & Distance between a transmitter and receiver pair & $\sim N\left(50,25\right)$\tabularnewline
\hline
$\tau$ & Modified FP NE check time & 60\tabularnewline
\hline
$\alpha$ & Modified FP step size & 0.5\tabularnewline
\hline
\end{tabular}
\end{table}

\section{Conclusion}

We used a utility design approach and constructed a non-cooperative
game such that all its pure NE have an asymptotically optimal weighted
sum-rate. By modifying the fictitious play algorithm for the wireless
environment, whose stable points are the pure NE of the designed game,
we designed a fully distributed channel allocation algorithm that
requires no communication between users. This means we have used game
theory as a distributed optimization tool.

Analyzing the performance of the designed game involved a novel probabilistic
analysis of the random pure NE of a random interference game in a
large network. We defined a bipartite graph of users and resources
that represents the $M$ best (or worst) REs of each user. We observed
that the random structure of our random NE is an ``almost'' perfect
matching in this random bipartite graph. By taking the number of users
($N$) to infinity, we were able to provide concentration results
on the existence (or non-existence) of pure NE in our random game.
This novel approach can be applied in other problems, by first identifying
the structure of the random NE.

The first game we analyzed is the naive non-cooperative game (Naive-FSIG),
where the utility function of each user is simply his achievable rate.
We showed that with strong enough interference it has $\varOmega\left(\left(N^{\mu}\right)!\right)$
(for all $\mu<1$) bad pure NE, where $N$ is the number of users.

Then we proposed a designed non-cooperative game formulation (M-FSIG)
whose utility is a slight modification of former, such that it is
greater than zero only for their $M$ best REs, with the same value
for those REs. We proved that asymptotically in $N$, all of its PNE
are almost a a perfect matching between users and $M$-best REs (or
an exact one) in our bipartite graph. This means that almost all users
get one of their $M$ best REs, interference-free.

In order to answer the question of how good is the $M$ best RE, we
analyzed the order statistics of the fading distribution. We defined
the family of exponentially-dominated tail distributions, that includes
many fading distributions (like Rayleigh fading), and showed that
for any such distribution the M-FSIG has a pure price of anarchy that
converges in probability to 1 as $N\rightarrow\infty$, in any interference
regime. Moreover, the M-FSIG exhibits $\varOmega\left(\left(N^{\mu}\right)!\right)$
(for all $\mu<1$) perfect matching pure NE that maintain max-min
fairness among the users.

We also proved that the asymptotic optimality of the M-FSIG holds
beyond the case of i.i.d. REs. All of our results on the M-FSIG hold
for $m$-dependent REs for each user, as appears in practice in OFDMA
systems.

For some fixed $N$ the introduced parameter $M$ can be chosen to
compromise between sum-rate and fairness. Due to the almost completely
orthogonal transmissions in equilibria, our allocation algorithm is
more suitable for the medium-strong interference regime.

We showed through simulations that our algorithm converges very fast
to the proven pure NE. The fast convergence enables frequent runs
of the algorithm in the network, which results in maintaining multi-user
diversity in a dynamic fading environment.

\appendices{}

\section{\label{Naive-FSIG Appendix}Proofs for the Naive-FSIG}

In this section we provide the proofs of the results in Section \ref{Naive-FSIG section}.
\begin{IEEEproof}[Proof of Proposition \ref{Worst Channels are Asymptotically Worthless}]
Let $\varepsilon>0$. Due to the i.i.d. assumption, the number $N_{\varepsilon,n}$
of r.v. from $\left|h_{n,n,1}\right|,\,...,\,\left|h_{n,n,N}\right|$
that are smaller than $\varepsilon$ has a binomial distribution with
$p_{n}=\Pr\left(\left|h_{n,n,1}\right|<\varepsilon\right)>0$. We
use the Chernoff-Hoeffding Theorem \cite{Hoeffding1963} as a tail
bound for $\frac{M_{N}}{N}<p_{n}$. By the assumption on $M_{N}$,
$\frac{M_{N}}{N}<p_{n}$ holds for all $N>N_{1}$ for some large enough
$N_{1}$, and so
\begin{equation}
\Pr\left(N_{\varepsilon,n}\leq M_{N}\right)\leq\exp\left(-ND\left(\frac{M_{N}}{N}\Vert p_{n}\right)\right)\label{eq:4}
\end{equation}
where $D(q\Vert p)=q\ln\frac{q}{p}+\left(1-q\right)\ln\frac{1-q}{1-p}$
, and in our case

\begin{equation}
D\left(\frac{M_{N}}{N}\Vert p_{n}\right)=\frac{M_{N}}{N}\ln\frac{M_{N}}{Np_{n}}+\left(1-\frac{M_{N}}{N}\right)\ln\frac{1-\frac{M_{N}}{N}}{1-p_{n}}\label{eq:5}
\end{equation}
for which we get by using $\underset{N\rightarrow\infty}{\lim}\frac{M_{N}}{N}=0$
that
\begin{multline}
\underset{N\rightarrow\infty}{\lim}D\left(\frac{M_{N}}{N}\Vert p_{n}\right)=-\underset{N\rightarrow\infty}{\lim}\frac{\ln\frac{N}{M_{N}}}{\frac{N}{M_{N}}}-\ln p_{n}\underset{N\rightarrow\infty}{\lim}\frac{M_{N}}{N}+\ln\left(\frac{1}{1-p_{n}}\right)\underset{N\rightarrow\infty}{\lim}\left(1-\frac{M_{N}}{N}\right)+\\
\underset{N\rightarrow\infty}{\lim}\left(1-\frac{M_{N}}{N}\right)\ln\left(1-\frac{M_{N}}{N}\right)=\ln\left(\frac{1}{1-p_{n}}\right).\label{eq:6}
\end{multline}
So for large enough $N$ the inequality $D\left(\frac{M_{N}}{N}\Vert p_{n}\right)\geq\ln\left(\frac{1}{1-p_{n}}\right)-\ln\left(\frac{1}{1-p_{n}^{2}}\right)$
holds; hence we get the following upper bound
\begin{equation}
\Pr\left(N_{\varepsilon,n}\leq M_{N}\right)\leq\exp\left(-ND\left(\frac{M_{N}}{N}\Vert p_{n}\right)\right)\leq\left(1-p_{n}\right)^{N}\left(\frac{1}{1-p_{n}^{2}}\right)^{N}=\left(\frac{1}{1+p_{n}}\right)^{N}\rightarrow0.\label{eq:7}
\end{equation}
Clearly, if there are at least $M_{N}$ successes then the $M_{N}$
smallest variables among $\left|h_{n,n,1}\right|,\,...,\,\left|h_{n,n,N}\right|$
are smaller than $\varepsilon$. Consequently, using the union bound
we get
\begin{equation}
\underset{N\rightarrow\infty}{\lim}\Pr\left(\underset{n}{\max\,}\left|h_{n,n,(m)}\right|>\varepsilon\right)=\underset{N\rightarrow\infty}{\lim}\Pr\left(\bigcup_{n=1}^{N}\left\{ \left|h_{n,n,(m)}\right|>\varepsilon\right\} \right)\leq\underset{N\rightarrow\infty}{\lim}\sum_{n=1}^{N}\frac{1}{\left(1+p_{n}\right)^{N}}=0\label{eq:8}
\end{equation}
for each $m\leq M_{N}$ and each $\varepsilon>0$, and we reach our
conclusion.
\end{IEEEproof}

\section{Exponentially-Dominated Tail Distributions\label{sec:Exponentially-Dominated-Tail-Dis}}

Our order statistics analysis provided in Appendix \ref{M-FSIG Appendix}
is valid for a broad family of fading distributions, named exponentially-dominated
tail distributions (see Definition \ref{Exponentially-Dominated Tail}).
In this appendix we develop the properties of these distributions
that are essential for our results.

The Rayleigh, Rician, m-Nakagami and Normal distributions all have
an exponentially-dominated tail. To ease the verification that a certain
distribution has an exponentially-dominated tail, we provide the following
lemma. The conditions of this lemma are easier to check than Definition
\ref{Exponentially-Dominated Tail} where the PDF has a more convenient
form than the CDF, or when a certain power of the original random
variable has a more convenient distribution.
\begin{lem}
Let $X$ be a positive random variable with a CDF $F_{X}$ and a PDF
$f_{X}$.

\begin{enumerate}
\item If for some $\alpha,\gamma,\lambda>0$ and $\beta\in\mathbb{R}$,
$\underset{x\rightarrow\infty}{\lim}\frac{f_{X}\left(x\right)}{\alpha\gamma\lambda x^{\beta+\gamma-1}e^{-\lambda x^{\gamma}}}=1$
holds then $X$ has an exponentially-dominated tail distribution.
\item For any $d>0$, if $X$ has an exponentially-dominated tail distribution
then so does $X^{d}$.
\end{enumerate}
\end{lem}
\begin{IEEEproof}
The first part follows from l'H�pital's rule
\begin{equation}
\underset{x\rightarrow\infty}{\lim}\frac{1-F_{X}\left(x\right)}{\alpha x^{\beta}e^{-\lambda x^{\gamma}}}=\underset{x\rightarrow\infty}{\lim}\frac{-f_{X}\left(x\right)}{-\alpha\gamma\lambda x^{\beta+\gamma-1}e^{-\lambda x^{\gamma}}+\alpha\beta x^{\beta-1}e^{-\lambda x^{\gamma}}}=\underset{x\rightarrow\infty}{\lim}\frac{\frac{f_{X}\left(x\right)}{\alpha\gamma\lambda x^{\beta+\gamma-1}e^{-\lambda x^{\gamma}}}}{1-\frac{\beta}{\gamma\lambda x^{\gamma}}}=1.\label{eq:17-4}
\end{equation}
For the second part, note that if $Y=X^{d}$ then $F_{Y}\left(y\right)=F_{X}\left(y^{1/d}\right)$;
hence $\underset{y\rightarrow\infty}{\lim}\frac{1-F_{Y}\left(y\right)}{\alpha y^{\frac{\beta}{d}}e^{-\lambda y^{\frac{\gamma}{d}}}}=\underset{y\rightarrow\infty}{\lim}\frac{1-F_{X}\left(y^{1/d}\right)}{\alpha\left(y^{1/d}\right)^{\beta}e^{-\lambda\left(y^{1/d}\right)^{\gamma}}}=1$.
\end{IEEEproof}
Note that the second part of the above lemma allows us to choose our
$X_{n,k}$ variable in Section V.I as either $\frac{1}{a_{n}}|h_{n,n,k}|^{2}$
or $\frac{1}{\sqrt{a_{n}}}$$|h_{n,n,k}|$, where $a_{n}=E\left\{ |h_{n,n,k}|^{2}\right\} $
for all $k$.

Due to our interest in the intermediate statistics compared to the
extreme statistics, the desired properties of exponentially-dominated
tail distributions will be expressed by the quantile function. The
following definition will simplify notations.
\begin{defn}
\label{def:Tail Quantile}Define the tail quantile function as $\bar{q}_{X}\left(p\right)=F_{X}^{-1}\left(1-p\right)=\min\left\{ x\,|\,F_{X}\left(x\right)\geq1-p\right\} $.
\end{defn}
The next proposition lists important properties of exponentially-dominated
tail distributions we will need for our proofs. These properties are
regrading the quantile function of an exponentially-dominated tail
distribution, and its intermediate to extreme statistics ratio.
\begin{prop}
\label{Eponentially-Dominated Tail Distribution Properites}Let $X$
be a random variable with a tail quantile function $\bar{q}_{X}\left(p\right)$
. If $X$ has an exponentially-dominated tail distribution with parameters
$\alpha,\gamma,\lambda>0$ and $\beta\in\mathbb{R}$, then

\begin{enumerate}
\item $\underset{p\rightarrow0}{\lim}\frac{\bar{q}_{X}\left(p\right)}{\left[\frac{1}{\lambda}\ln\left(\alpha\frac{1}{p}\right)\right]^{1/\gamma}}=1$.
\item If $\underset{K\rightarrow\infty}{\lim}\frac{M_{K}}{\ln^{\mu}K}=0$
for some $\mu>0$ then $\underset{K\rightarrow\infty}{\lim}\frac{\bar{q}_{X}\left(\frac{M_{K}}{K}\right)}{\bar{q}_{X}\left(\frac{1}{K}\right)}=1$.
\item If $\underset{K\rightarrow\infty}{\lim}\frac{M_{K}}{K^{\mu}}=0$ for
some $\mu<1$ then $\underset{K\rightarrow\infty}{\lim}\frac{\bar{q}_{X}\left(\frac{M_{K}}{K}\right)}{\bar{q}_{X}\left(\frac{1}{K^{2}}\right)}>0$.
\end{enumerate}
\end{prop}
\begin{IEEEproof}
Define $g_{1}\left(x\right)=\alpha e^{-\lambda\left(x^{\gamma}+2x^{\gamma/2}\right)}$,
$g_{2}\left(x\right)=\alpha e^{-\lambda\left(x^{\gamma}-2x^{\gamma/2}\right)}$.
Let $\varepsilon>0$. For some large enough $x_{1}>0$, the inequality
\begin{equation}
\frac{1}{1-\varepsilon}\alpha e^{-\lambda\left(x^{\gamma}+2x^{\gamma/2}\right)}\leq\alpha x^{\beta}e^{-\lambda x^{\gamma}}\leq\frac{1}{1+\varepsilon}\alpha e^{-\lambda\left(x^{\gamma}-2x^{\gamma/2}\right)}\label{eq:17-1}
\end{equation}
holds for all $x>x_{1}$. Due to the exponentially-dominated tail,
for some large enough $x_{2}>0$, the inequality
\begin{equation}
\left(1-\varepsilon\right)\alpha x^{\beta}e^{-\lambda x^{\gamma}}\leq1-F_{X}\left(x\right)\leq\left(1+\varepsilon\right)\alpha x^{\beta}e^{-\lambda x^{\gamma}}\label{eq:17-2}
\end{equation}
holds for all $x>x_{2}$. Combining \eqref{eq:17-1} and \eqref{eq:17-2}
we conclude that for all $x>\max\left\{ x_{1},x_{2}\right\} $ the
following inequality holds
\begin{equation}
g_{1}\left(x\right)=\alpha e^{-\lambda\left(x^{\gamma}+2x^{\gamma/2}\right)}\leq1-F_{X}\left(x\right)\leq\alpha e^{-\lambda\left(x^{\gamma}-2x^{\gamma/2}\right)}=g_{2}\left(x\right).\label{eq:17-3}
\end{equation}
For small enough $p$, the tail quantile function $\bar{q}_{X}\left(p\right)$
is large enough, i.e., $\bar{q}_{X}\left(p\right)>\max\left\{ x_{1},x_{2}\right\} $
and from \eqref{eq:17-3} we get
\begin{equation}
\begin{array}{c}
g_{1}\left(\bar{q}_{X}\left(p\right)\right)\leq p=1-F_{X}\left(\bar{q}_{X}\left(p\right)\right)\leq g_{2}\left(\bar{q}_{X}\left(p\right)\right)\end{array}.\label{eq:18}
\end{equation}
It is easy to verify that for $p\leq\alpha$, such that $\ln\left(\frac{\alpha}{p}\right)\geq0$,
we have
\begin{equation}
\begin{array}{c}
g_{1}^{-1}\left(p\right)=\left[\sqrt{\frac{1}{\lambda}\ln\left(\frac{\alpha}{p}\right)+1}-1\right]^{2/\gamma}\\
\\
g_{2}^{-1}\left(p\right)=\left[\sqrt{\frac{1}{\lambda}\ln\left(\frac{\alpha}{p}\right)+1}+1\right]^{2/\gamma}
\end{array}.\label{eq:19}
\end{equation}
Hence by invoking $g_{1}^{-1}\left(p\right),g_{2}^{-1}\left(p\right)$,
which are monotonically decreasing, on the first and second inequalities
in \eqref{eq:18} respectively, we get that for small enough $p$
the following holds
\begin{equation}
\left[\sqrt{\frac{1}{\lambda}\ln\left(\frac{\alpha}{p}\right)+1}-1\right]^{2/\gamma}\leq\bar{q}_{X}\left(p\right)\leq\left[\sqrt{\frac{1}{\lambda}\ln\left(\frac{\alpha}{p}\right)+1}+1\right]^{2/\gamma}\label{eq:20}
\end{equation}
which leads to
\begin{equation}
\underset{p\rightarrow0}{\lim}\frac{\bar{q}_{X}\left(p\right)}{\left[\frac{1}{\lambda}\ln\left(\frac{\alpha}{p}\right)\right]^{1/\gamma}}\geq\underset{p\rightarrow0}{\lim}\frac{\left[\sqrt{\frac{1}{\lambda}\ln\left(\frac{\alpha}{p}\right)+1}-1\right]^{2/\gamma}}{\left[\frac{1}{\lambda}\ln\left(\frac{\alpha}{p}\right)\right]^{1/\gamma}}=1\label{eq:21}
\end{equation}
and
\begin{equation}
\underset{p\rightarrow0}{\lim}\frac{\bar{q}_{X}\left(p\right)}{\left[\frac{1}{\lambda}\ln\left(\frac{\alpha}{p}\right)\right]^{1/\gamma}}\leq\underset{p\rightarrow0}{\lim}\frac{\left[\sqrt{\frac{1}{\lambda}\ln\left(\frac{\alpha}{p}\right)+1}+1\right]^{2/\gamma}}{\left[\frac{1}{\lambda}\ln\left(\frac{\alpha}{p}\right)\right]^{1/\gamma}}=1\label{eq:23-1}
\end{equation}
where both limits stem from $\left[\frac{1}{\lambda}\ln\left(\frac{\alpha}{p}\right)\right]^{1/\gamma}$
being the dominant term in $\left[\sqrt{\frac{1}{\lambda}\ln\left(\frac{\alpha}{p}\right)+1}\pm1\right]^{2/\gamma}$.
Together \eqref{eq:21} and \eqref{eq:23-1} yield $\underset{p\rightarrow0}{\lim}\frac{\bar{q}_{X}\left(p\right)}{\left[\frac{1}{\lambda}\ln\left(\frac{\alpha}{p}\right)\right]^{1/\gamma}}=1$.
Now let $M_{K}$ be a sequence such that $\underset{K\rightarrow\infty}{\lim}\frac{M_{K}}{\ln^{\mu}K}=0$
for some $\mu>0$ and obtain
\begin{equation}
\underset{K\rightarrow\infty}{\lim}\frac{\bar{q}_{X}\left(\frac{M_{K}}{K}\right)}{\bar{q}_{X}\left(\frac{1}{K}\right)}\underset{(a)}{=}\underset{K\rightarrow\infty}{\lim}\frac{\bar{q}_{X}\left(\frac{M_{K}}{K}\right)}{\left[\frac{1}{\lambda}\ln\left(\alpha\frac{K}{M_{K}}\right)\right]^{1/\gamma}}\underset{K\rightarrow\infty}{\lim}\frac{\left[\frac{1}{\lambda}\ln\left(\alpha K\right)\right]^{1/\gamma}}{\bar{q}_{X}\left(\frac{1}{K}\right)}\underset{(b)}{=}1\label{eq:22}
\end{equation}
where (a) follows from $\underset{K\rightarrow\infty}{\lim}\frac{\left[\frac{1}{\lambda}\ln\left(\alpha K\right)\right]^{1/\gamma}}{\left[\frac{1}{\lambda}\ln\left(\alpha\frac{K}{M_{K}}\right)\right]^{1/\gamma}}=1$,
which is true due to $\underset{K\rightarrow\infty}{\lim}\frac{M_{K}}{\ln^{\mu}K}=0$
for some $\mu>0$, and (b) from Part 1. Now let $M_{K}$ be a sequence
such that $\underset{K\rightarrow\infty}{\lim}\frac{M_{K}}{K^{\mu}}=0$
for some $\mu<1$ and obtain

\begin{equation}
\underset{K\rightarrow\infty}{\lim}\frac{\bar{q}_{X}\left(\frac{M_{K}}{K}\right)}{\bar{q}_{X}\left(\frac{1}{K^{2}}\right)}\underset{(a)}{=}\underset{K\rightarrow\infty}{\lim}\frac{\left[\frac{1}{\lambda}\ln\left(\alpha\frac{K}{M_{K}}\right)\right]^{1/\gamma}}{\left[\frac{1}{\lambda}\ln\left(\alpha K^{2}\right)\right]^{1/\gamma}}=\left(\frac{1}{2}\right)^{1/\gamma}\underset{K\rightarrow\infty}{\lim}\left(\frac{\ln\left(\alpha\frac{K}{M_{K}}\right)}{\ln\left(\sqrt{\alpha}K\right)}\right)^{1/\gamma}\underset{(b)}{>}0\label{eq:23}
\end{equation}
where (a) follows from $\underset{K\rightarrow\infty}{\lim}\frac{\bar{q}_{X}\left(\frac{M_{K}}{K}\right)}{\left[\frac{1}{\lambda}\ln\left(\alpha\frac{K}{M_{K}}\right)\right]^{1/\gamma}}=1$
and $\underset{K\rightarrow\infty}{\lim}\frac{\bar{q}_{X}\left(\frac{1}{K^{2}}\right)}{\left[\frac{1}{\lambda}\ln\left(\alpha K^{2}\right)\right]^{1/\gamma}}=1$,
which we proved in Part 1, and (b) from the assumption that $\underset{K\rightarrow\infty}{\lim}\frac{M_{K}}{K^{\mu}}=0$
for some $\mu<1$.
\end{IEEEproof}
It should be emphasized that, technically speaking, the only result
in this paper that actually requires the exponentially-dominated tail
assumption is Proposition \ref{Eponentially-Dominated Tail Distribution Properites}
above. Also note that Parts 2 and 3 of this Proposition follow from
Part 1 directly. Therefore, all our order statistics analysis automatically
applies to any distribution $F_{X}\left(x\right)$ with a tail quantile
function $\bar{q}_{X}\left(p\right)$ such that $\underset{p\rightarrow0}{\lim}\frac{\bar{q}_{X}\left(p\right)}{\left[\frac{1}{\lambda}\ln\left(\alpha\frac{1}{p}\right)\right]^{1/\gamma}}=1$.
The question of how broader this class than the family of exponentially-dominated
tail distributions remains open.

\section{\label{M-FSIG Appendix}Proofs for the M-FSIG}

In this section we provide the proofs of the results in Section \ref{M-FSIG Section}.

\subsection{Proof of Theorem \ref{Non-Existence of Bad Equilibria }}
\begin{IEEEproof}
Denote the number of shared REs by $K_{c}$. Denote the number of
sharing users by $N_{c}$. In each shared RE, every user except one
contributes one empty RE to the total number of empty REs $E$, so
$E=N_{c}-K_{c}$. Every shared RE must contain at least two users;
hence $\frac{N_{c}}{2}\leq E\leq N_{c}$. In a PNE for the M-FSIG,
no empty RE can be one of the $M$- best REs of one of the sharing
users.

Let $\mathcal{E}$ be a set of empty REs with cardinality $E$ and
$\mathcal{S}$ be a set of sharing users with cardinality $N_{c}$.
We want to upper bound the probability that user $n$ will not have
any of the $E$ empty REs in $\mathcal{E}$ in his $M$-best RE list.
For i.i.d. channel gains this can be done as follows
\begin{multline}
\Pr\left(\mathcal{M}_{n}\cap\mathcal{E}=\emptyset\right)\underset{(a)}{=}\frac{\left(\begin{array}{c}
K-E\\
M
\end{array}\right)}{\left(\begin{array}{c}
K\\
M
\end{array}\right)}=\frac{(K-E)!(K-M)!}{K!(K-E-M)!}=\prod_{i=1}^{E}\frac{K-E-M+i}{K-E+i}=\\
\prod_{i=1}^{E}\left(1-\frac{M}{K-E+i}\right)\underset{(b)}{\leq}\left(1-\frac{M}{K}\right)^{E}\underset{(c)}{\leq}e^{-M\frac{E}{K}}\underset{(d)}{\leq}e^{-M\frac{N_{c}}{2K}}\label{eq:1-2}
\end{multline}
where (a) follows from $h_{n,n,1},\,...,\,h_{n,n,k}$ being i.i.d.,
(b) follows from $1-\frac{M}{K-E+i}\leq1-\frac{M}{K}$ for $i=1,...,E$,
(c) from $\left(1-\frac{M}{K}\right)^{K}\leq e^{-M}$ and (d) from
$E\geq\frac{N_{c}}{2}$. For the $m$-dependent case we have instead,
from Lemma \ref{lem:m-dependent bounds} that
\begin{multline}
\Pr\left(\mathcal{M}_{n}\cap\mathcal{E}=\emptyset\right)\leq\left(m+1\right)e^{-\frac{M}{e\left(m+1\right)}\frac{N-m-1}{N+m+1}}+\left(1-\frac{M}{e\left(N+m+1\right)}\right)^{\frac{E}{m+1}}\underset{(a)}{\leq}\\
\left(m+1\right)e^{-\frac{M}{e\left(m+1\right)}\frac{N-m-1}{N+m+1}}+e^{-\frac{M}{e\left(m+1\right)}\frac{E}{N+m+1}}\underset{(b)}{\leq}\left(m+2\right)e^{-\frac{M}{e\left(m+1\right)}\frac{E}{N+m+1}}\label{eq:1-3}
\end{multline}
where (a) is from $\left(1-\frac{x}{L}\right)^{L}\leq e^{-x}$ which
is true for each $x<L$ where $L$ is a positive integer, and (b)
by comparing the exponents for $E\leq N-m-1$. Denote by $\mathcal{E_{\mathbf{a}}}$
and $\mathcal{S_{\mathbf{a}}}$ the set of empty REs and the set of
sharing users, respectively, of the strategy profile $\mathbf{a}$.
We get, using \eqref{eq:1-3}, that
\begin{multline}
\Pr\left(\left\{ \mathbf{a}\,|\,\mathcal{\mathcal{E}\,\subseteq E_{\mathbf{a}}},\,\mathcal{S\subseteq S_{\mathbf{a}}}\right\} \cap\mathcal{P}_{e}\neq\emptyset\right)\underset{(a)}{\leq}\\
\Pr\left(\bigcap_{n\in\mathcal{S}}\left\{ \mathcal{M}_{n}\cap\mathcal{E}=\emptyset\right\} \right)\underset{(b)}{=}\bigcap_{n\in\mathcal{S}}\Pr\left(\mathcal{M}_{n}\cap\mathcal{E}=\emptyset\right)\underset{(c)}{\leq}\left(m+2\right)^{N_{c}}e^{-\frac{M}{2e\left(m+1\right)}\frac{N_{c}^{2}}{N+m+1}}\label{eq:2-2}
\end{multline}
where (a) is because $\left\{ \mathbf{a}\,|\,\mathcal{\mathcal{E}\,\subseteq E_{\mathbf{a}}},\,\mathcal{S\subseteq S_{\mathbf{a}}}\right\} \cap\mathcal{P}_{e}\neq\emptyset$
implies $\bigcap_{n\in\mathcal{S}}\left\{ \mathcal{M}_{n}\cap\mathcal{E}=\emptyset\right\} $,
(b) is due to the independence of users' preferences, and (c) from
\eqref{eq:1-3}. This means that the set of all the strategy profiles
with empty REs and sharing users sets that contain $\mathcal{E}$
and $\mathcal{S}$, respectively, have a vanishing probability to
contain a PNE, if $N_{c}=rN$ for some $0<r<1$. Now we want to show
that even$\underset{\mathbf{a^{*}}\in\mathcal{P}_{e}}{\max}\frac{N_{c}\left(\mathbf{a^{*}}\right)}{N}\rightarrow0$
as $N\rightarrow\infty$. We want to find a probabilistic upper bound
$\tilde{N_{c}}$ for the number of sharing users that a PNE may contain.
We use the union bound over all the choices of $\tilde{N_{c}}$ sharing
users and the set of $\tilde{E}=\frac{\tilde{N_{c}}}{2}$ necessarily
existing empty REs, to get from \eqref{eq:2-2} that
\begin{multline}
\Pr\left(\underset{\mathbf{a^{*}}\in\mathcal{P}_{e}}{\max}N_{c}\left(\mathbf{a^{*}}\right)\geq\tilde{N_{c}}\right)\underset{(a)}{=}\Pr\left(\left\{ \mathbf{a}\,|\,\left|\mathcal{S_{\mathbf{a}}}\right|\geq\tilde{N_{c}}\,,\,\left|\mathcal{E_{\mathbf{a}}}\right|\geq\tilde{E}\right\} \cap\mathcal{P}_{e}\neq\emptyset\right)\underset{(b)}{=}\\
\Pr\left(\bigcup_{\mathcal{\mathit{\mathcal{S},\mathcal{E}\in\Lambda}}}\biggl\{\left\{ \mathbf{a}\,|\,\mathcal{\mathcal{E}\,\subseteq E_{\mathbf{a}}},\,\mathcal{S\subseteq S_{\mathbf{a}}}\right\} \cap\mathcal{P}_{e}\neq\emptyset\biggr\}\right)\underset{(c)}{\leq}\left(\begin{array}{c}
N\\
\tilde{N_{c}}
\end{array}\right)\left(\begin{array}{c}
K\\
\tilde{E}
\end{array}\right)\left(m+2\right)^{\tilde{N_{c}}}e^{-\frac{M}{2e\left(m+1\right)}\frac{\tilde{N_{c}}^{2}}{N+m+1}}.\label{eq:3-2}
\end{multline}
The event $\underset{\mathbf{a^{*}}\in\mathcal{P}_{e}}{\max}N_{c}\left(\mathbf{a^{*}}\right)\geq\tilde{N_{c}}$
occurs if and only if some PNE has at least $\tilde{N_{c}}$ sharing
users. This is used in (a). Equality (b) follows from going over all
the options in $\Lambda=\left\{ \mathcal{S},\mathcal{E}\,|\,\left|\mathcal{S}\right|=\tilde{N_{c}},\left|\mathcal{E}\right|=\tilde{E}\right\} $,
and (c) is the union bound. The inequality in \eqref{eq:3-2} bounds
the probability that there exists a PNE with at least $\tilde{N_{c}}$
sharing users. Now choose $\tilde{N_{c}}=rN$ for some $0<r\leq\frac{1}{2}$
, so $\left(\begin{array}{c}
K\\
\tilde{E}
\end{array}\right)\leq\left(\begin{array}{c}
N\\
\tilde{N_{c}}
\end{array}\right)$. We can use the bound $\left(\begin{array}{c}
N\\
rN
\end{array}\right)\leq\sqrt{\frac{1}{2\pi r\left(1-r\right)N}}2^{Nh_{2}\left(r\right)}$ \cite{Cover2012} where $h_{2}(r)$ is the binary entropy of $r$
and obtain from \eqref{eq:3-2} that
\begin{multline}
\Pr\left(\underset{\mathbf{a^{*}}\in\mathcal{P}_{e}}{\max}N_{c}\left(\mathbf{a^{*}}\right)\geq rN\right)\leq\frac{\left(m+2\right)^{rN}}{2\pi r\left(1-r\right)N}4^{Nh_{2}(r)}e^{-\frac{M}{2e\left(m+1\right)}\frac{N}{N+m+1}r^{2}N}\underset{(a)}{\leq}\\
\frac{1}{2\pi r\left(1-r\right)N}\left(\frac{\left(m+2\right)^{r}4^{h_{2}(r)}}{e^{\frac{r^{2}}{6\left(m+1\right)}M}}\right)^{N}\underset{(b)}{\leq}\frac{1}{2\pi r\left(1-r\right)N}\left(\frac{\left(m+2\right)^{r}4^{h_{2}(r)}}{N^{r^{2}\frac{\varepsilon+e}{6}}}\right)^{N}\label{eq:16-1-1}
\end{multline}
where (a) is due to $\frac{N}{N+m+1}\geq\frac{2e}{6}$ for large enough
$N$ and (b) follows from $M\geq\left(e+\varepsilon\right)\left(m+1\right)\ln(N)$.
In conclusion we get that $\underset{N\rightarrow\infty}{\lim}\Pr\left(\underset{\mathbf{a^{*}}\in\mathcal{P}_{e}}{\max}\frac{N_{c}\left(\mathbf{a^{*}}\right)}{N}<r\right)=1$
for all $r>0$ and hence $\underset{\mathbf{a^{*}}\in\mathcal{P}_{e}}{\max}\frac{N_{c}\left(\mathbf{a^{*}}\right)}{N}\rightarrow0$
in probability as $N\rightarrow\infty$.

Note that substituting $m=0$ in \eqref{eq:16-1-1} yields a slightly
looser bound than we can get with a direct analysis of i.i.d. channel
gains. With i.i.d. channel gains, the factor $\left(m+2\right)^{r}$
disappears and $N^{r^{2}\frac{\varepsilon+e}{6}}$ is replaced by
$N^{r^{2}\frac{\varepsilon+e}{2}}$. This is easily verified using
\eqref{eq:1-2} instead of \eqref{eq:1-3}.
\end{IEEEproof}

\subsection{Proofs of Theorem \ref{non-log Asymptotic Optimality} and Theorem
\ref{Max-Min Fairness}}

In order to prove Theorem \ref{non-log Asymptotic Optimality}, we
need the following two lemmas that provide probabilistic bounds on
the relevant random variables.

The first is a probabilistic upper bound for the extreme statistics
of an exponentially-dominated tail distribution. Note that this lemma
assumes nothing regrading the independence of the series of random
variables.
\begin{lem}
\label{Extreme Statistics Lemma}Let $\left\{ X_{n,k}\right\} $ be
identically distributed unbounded random variables with exponentially-dominated
tail distribution $F_{X}$ with parameters $\alpha,\gamma,\lambda>0$
and $\beta\in\mathbb{R}$. If $U_{K}=\left(\frac{\ln\left(K\right)+\sqrt{\ln\left(K\right)}}{\lambda}\right)^{1/\gamma}$
then

\begin{enumerate}
\item $\underset{K\rightarrow\infty}{\lim}\Pr\left(X_{n,(K)}\leq U_{K}\right)=1$
for each $n\in\mathcal{N}$.
\item $\underset{N\rightarrow\infty}{\lim}\Pr\left(\underset{n}{\max\,}X_{n,(K)}\leq U_{N^{2}}\right)=1$.
\end{enumerate}
\end{lem}
\begin{IEEEproof}
From the union bound and the identical distributions of $X_{n,1},...,X_{n,K}$
we obtain
\begin{equation}
\Pr\left(X_{n,(K)}\geq U_{K}\right)=\Pr\left(\bigcup_{k=1}^{K}\left\{ X_{n,k}\geq U_{K}\right\} \right)\leq K\Pr\left(X_{n,1}\geq U_{K}\right).\label{eq:35}
\end{equation}
We get that for $U_{K}=\left(\frac{\ln\left(K\right)+\sqrt{\ln\left(K\right)}}{\lambda}\right)^{1/\gamma}$
\begin{multline}
\underset{K\rightarrow\infty}{\lim}K\Pr\left(X_{n,1}\geq U_{K}\right)=\underset{K\rightarrow\infty}{\lim}\frac{\Pr\left(X_{n,1}\geq U_{K}\right)}{\alpha U_{K}^{\beta}e^{-\lambda U_{K}^{\gamma}}}\underset{K\rightarrow\infty}{\lim}\frac{\alpha U_{K}^{\beta}K}{\exp\left(\ln\left(K\right)+\sqrt{\ln\left(K\right)}\right)}\underset{(a)}{=}\\
\frac{\alpha}{\lambda^{\beta/\gamma}}\underset{K\rightarrow\infty}{\lim}\left(\ln\left(K\right)+\sqrt{\ln\left(K\right)}\right)^{\beta/\gamma}\exp\left(-\sqrt{\ln\left(K\right)}\right)=0\label{eq:36}
\end{multline}
where (a) is from the definition of an exponentially-dominated tail.
This proves the first part of the lemma. For the second part consider
all the variables $\left\{ X_{n,k}\right\} $, as $NK=N^{2}$ identically
distributed samples from the parent distribution $F_{X}\left(x\right)$,
and define $X_{(N^{2})}=\underset{n}{\max\,}X_{n,(K)}$. By substituting
$K=N^{2}$ in the derivation above the result readily follows.
\end{IEEEproof}
Next we prove the probabilistic lower bound for the intermediate statistics
of an exponentially-dominated tail distribution. Note that the next
lemma holds for m-dependent sequences of random variables.
\begin{lem}
\label{Intermediate Statistics Lemma}Let $\left\{ X_{n,k}\right\} $
be identically distributed unbounded random variables with an exponentially-dominated
tail distribution $F_{X}$. Assume that $\left\{ X_{n,k}\right\} _{k}$
is $m$-dependent for each $n$. Let $M_{K}$ be a sequence such that
$\underset{K\rightarrow\infty}{\lim}\frac{M_{K}}{K}=0$ and $\underset{K\rightarrow\infty}{\lim}M_{K}=\infty$
and define $L_{K}=F_{X}^{-1}\left(1-\frac{e^{2}M_{K}}{K}\right)$.
If $M_{K}\geq\left(m+1\right)(1+\varepsilon)\ln K$ for some $\varepsilon>0$
then

\begin{enumerate}
\item For large enough $K$ we have $\Pr\left(X_{n,(K-M_{K}+1)}\geq L_{K}\right)\geq1-\frac{2\left(m+1\right)}{K^{4(1+\varepsilon)}}$
for each $n\in\mathcal{N}$.
\item $\underset{K\rightarrow\infty}{\lim}\Pr\left(\underset{n}{\min}\,X_{n,(K-M_{K}+1)}\geq L_{K}\right)=1$.
\end{enumerate}
\end{lem}
\begin{IEEEproof}
Let $L_{K}=\bar{q}_{X}\left(\frac{e^{2}M_{K}}{K}\right)$. By invoking
$F_{K}\left(x\right)=\sum_{k=1}^{K}I\left(X_{n,k}\leq x\right)$ on
both sides of $X_{n,(K-M_{K}+1)}\leq L_{K}$ we get
\begin{equation}
\Pr\left(X_{n,(K-M_{K}+1)}\leq L_{K}\right)=\Pr\left(K-M_{K}+1\leq\sum_{k=1}^{K}I\left(X_{n,k}\leq L_{K}\right)\right)\underset{(a)}{=}\Pr\left(\sum_{k=1}^{K}I\left(X_{n,k}>L_{K}\right)<M_{K}\right)\label{eq:37}
\end{equation}
where (a) follows by $K-\sum_{k=1}^{K}I\left(X_{n,k}\leq L_{K}\right)=\sum_{k=1}^{K}I\left(X_{n,k}>L_{K}\right)$.
Now we divide $X_{n,1},...,X_{n,K}$ into $m+1$ disjoint sets $\left\{ X_{n,1},X_{n,2+m},...\right\} ,...,\left\{ X_{n,m+1},X_{n,2+2m},...\right\} $
and define $Y_{n,i,j}=X_{n,j+\left(m+1\right)i}$ for $j=1,..,m+1$
and all $i$ such that $1\leq j+\left(m+1\right)i\leq K$ for some
$j$. Note that there are $\left\lfloor \frac{K}{m+1}\right\rfloor +1$
or $\left\lfloor \frac{K}{m+1}\right\rfloor $ elements in each of
these sets. By omitting the last element in the larger sets we obtain
\begin{equation}
\sum_{k=1}^{K}I\left(X_{n,k}>L_{K}\right)\geq\sum_{j=1}^{m+1}\sum_{i=0}^{\left\lfloor \frac{K}{m+1}\right\rfloor -1}I\left(X_{n,j+\left(m+1\right)i}>L_{K}\right)=\sum_{j=1}^{m+1}\sum_{i=0}^{\left\lfloor \frac{K}{m+1}\right\rfloor -1}I\left(Y_{n,i,j}>L_{K}\right).\label{eq:38}
\end{equation}
Because $X_{n,1},...,X_{n,K}$, are $m$-dependent, $Y_{n,0,j},...,Y_{n,\left\lfloor \frac{K}{m+1}\right\rfloor -1,j}$
are independent random variables for each $j$. For large enough $K$,
$\frac{\left(m+1\right)M_{K}}{K}<1$ and so
\begin{equation}
\Pr\left(\sum_{k=1}^{K}I\left(X_{n,k}>L_{K}\right)<M_{K}\right)=\Pr\left(\sum_{k=1}^{K}I\left(X_{n,k}>L_{K}\right)<M_{K}-\frac{\left(m+1\right)M_{K}}{K}\right)\label{eq:39}
\end{equation}
since $\sum_{k=1}^{K}I\left(X_{n,k}>L_{K}\right)$ is an integer.
Denote $t=\left(m+1\right)\frac{M_{K}}{K}\left\lfloor \frac{K}{m+1}\right\rfloor $
and note that $M_{K}-\frac{\left(m+1\right)M_{K}}{K}\leq t\leq M_{K}$.
By the union bound and the identical distributions of $\left\{ X_{n,k}\right\} $
we get, for large enough $K$, that

\begin{multline}
\Pr\left(\sum_{k=1}^{K}I\left(X_{n,k}>L_{K}\right)<M_{K}\right)\underset{(a)}{=}\Pr\left(\sum_{k=1}^{K}I\left(X_{n,k}>L_{K}\right)<t\right)\underset{(b)}{\leq}\Pr\left(\sum_{j=1}^{m+1}\sum_{i=0}^{\left\lfloor \frac{K}{m+1}\right\rfloor -1}I\left(Y_{n,i,j}>L_{K}\right)<t\right)\underset{(c)}{\leq}\\
\Pr\left(\bigcup_{j=1}^{m+1}\left\{ \sum_{i=0}^{\left\lfloor \frac{K}{m+1}\right\rfloor -1}I\left(Y_{n,i,j}\geq L_{K}\right)<\frac{t}{m+1}\right\} \right)\underset{(d)}{\leq}\left(m+1\right)\Pr\left(\sum_{i=0}^{\left\lfloor \frac{K}{m+1}\right\rfloor -1}I\left(Y_{n,i,1}>L_{K}\right)<\frac{t}{m+1}\right)\label{eq:40}
\end{multline}
where (a) follows from \eqref{eq:39}, (b) from \eqref{eq:38}, (c)
because some inner sum must be smaller than $\frac{t}{m+1}$ for the
total sum to be smaller than $t$ and (d) from the union bound. Define
the random variables $Z_{i}=I\left(Y_{n,i,1}\geq L_{K}\right)$ for
$i=0,..,\left\lfloor \frac{K}{m+1}\right\rfloor -1$. Due to the independence
of $Y_{n,0,j},...,Y_{n,\left\lfloor \frac{K}{m+1}\right\rfloor -1,j}$,
these variables form a Bernoulli process with success probability
$p=1-F_{X}\left(L_{K}\right)=\frac{e^{2}M_{K}}{K}$. Now we can apply
a concentration upper bound on the last term using a Chernoff bound
for a Binomially distributed variable. Because $p>\frac{M_{K}}{K}$
we obtain, after substituting $t=\left(m+1\right)\frac{M_{K}}{K}\left\lfloor \frac{K}{m+1}\right\rfloor $,
that

\begin{equation}
\Pr\left(\sum_{i=0}^{\left\lfloor \frac{K}{m+1}\right\rfloor -1}I\left(Y_{n,i,1}>L_{K}\right)<\frac{\left(m+1\right)\frac{M_{K}}{K}\left\lfloor \frac{K}{m+1}\right\rfloor }{m+1}\right)\leq\\
\exp\left(-\left\lfloor \frac{K}{m+1}\right\rfloor D\left(\frac{M_{K}}{K}||p\right)\right)\label{eq:41}
\end{equation}
and for large enough $K$
\begin{equation}
D\left(\frac{M_{K}}{K}||\frac{e^{2}M_{K}}{K}\right)=-\frac{M_{K}}{K}\ln\left(e^{2}\right)+\left(1-\frac{M_{K}}{K}\right)\ln\left(1+\frac{\left(e^{2}-1\right)\frac{M_{K}}{K}}{1-\frac{e^{2}M_{K}}{K}}\right)\underset{(a)}{\geq}-\frac{2M_{K}}{K}+\frac{1-\frac{M_{K}}{K}}{1-\frac{e^{2}M_{K}}{K}}\frac{6M_{K}}{K}\underset{(b)}{\geq}\frac{4M_{K}}{K}\label{eq:42}
\end{equation}
where (a) follows from $\ln\left(1+x\right)\geq\frac{6}{e^{2}-1}x$
for all $x\leq\frac{1}{100}$ together with $\underset{K\rightarrow\infty}{\lim}\frac{M_{K}}{K}=0$
and (b) follows from $\frac{1-\frac{M_{K}}{K}}{1-\frac{e^{2}M_{K}}{K}}\geq1$.
So for large enough $K$ the inequality $D\left(\frac{M_{K}}{K}||\frac{e^{2}M_{K}}{K}\right)\geq\frac{4M_{K}}{K}$
holds. If $M_{K}\geq\left(m+1\right)(1+\varepsilon)\ln\left(K\right)$
for some $\varepsilon>0$ then, we conclude from \eqref{eq:37}, \eqref{eq:40},
\eqref{eq:41} and \eqref{eq:42} that for large enough $K$

\begin{multline}
\Pr\left(X_{n,(K-M_{K}+1)}\leq L_{K}\right)\leq\left(m+1\right)\exp\left(-\left\lfloor \frac{K}{m+1}\right\rfloor D\left(\frac{M_{K}}{K}||\frac{e^{2}M_{K}}{K}\right)\right)\leq\\
\left(m+1\right)\exp\left(-4\left(\frac{M_{K}}{m+1}-\frac{M_{K}}{K}\right)\right)\leq\frac{2\left(m+1\right)}{K^{4(1+\varepsilon)}}\label{eq:43}
\end{multline}
which proves part one, and readily follows to part two by
\begin{multline}
\Pr\left(\underset{n}{\min}\,X_{n,(K-M_{K}+1)}\geq L_{K}\right)=\Pr\left(\bigcap_{n}\left\{ X_{n,(K-M_{K}+1)}\geq L_{K}\right\} \right)\underset{(a)}{\geq}\\
N\left(1-\frac{2\left(m+1\right)}{K^{4(1+\varepsilon)}}\right)-(N-1)\underset{(b)}{=}1-\frac{2\left(m+1\right)}{K^{3+4\varepsilon}}\underset{K\rightarrow\infty}{\rightarrow}1\label{eq:44}
\end{multline}
where (a) is from Fr�chet inequality and \eqref{eq:43} and (b) due
to $N=K$.
\end{IEEEproof}
Using the above lemmas, we can now prove both Theorem \ref{non-log Asymptotic Optimality}
and Theorem \ref{Max-Min Fairness}.
\begin{IEEEproof}[Proof of Theorem \ref{non-log Asymptotic Optimality}.]
Define $U_{K}=\left(\frac{\ln\left(K\right)+\sqrt{\ln\left(K\right)}}{\lambda}\right)^{1/\gamma}$
and $L_{K}=\overline{q}_{X}\left(\frac{e^{2}M_{K}}{K}\right)$ (see
Definition \ref{def:Tail Quantile}). We get the following inequality
\begin{multline}
\Pr\left(\frac{X_{(K-M_{K}+1)}}{X_{(K)}}\geq\frac{L_{K}}{U_{K}}\right)\underset{(a)}{\geq}\Pr\left(X_{(K-M_{K}+1)}\geq L_{K}\,,\,X_{(K)}\leq U_{K}\right)\underset{(b)}{\geq}\\
1-\Pr\left(X_{(K-M_{K}+1)}\leq L_{K}\right)-\Pr\left(X_{(K)}\geq U_{K}\right)\label{eq:24}
\end{multline}
where (a) is due to $\left\{ X_{(K-M_{K}+1)}\geq L_{K},X_{(K)}\leq U_{K}\right\} \subseteq\left\{ \frac{X_{(K-M_{K}+1)}}{X_{(K)}}\geq\frac{L_{K}}{U_{K}}\right\} $
and (b) is the union bound. From Lemma \ref{Extreme Statistics Lemma}
Part 1 and Lemma \ref{Intermediate Statistics Lemma} Part 1 we get
$\underset{K\rightarrow\infty}{\lim}\Pr\left(\frac{X_{(K-M_{K}+1)}}{X_{(K)}}\geq\frac{L_{K}}{U_{K}}\right)=1$.
From Proposition \ref{Eponentially-Dominated Tail Distribution Properites}
(Parts 1 and 2)
\begin{equation}
\underset{K\rightarrow\infty}{\lim}\frac{L_{K}}{U_{K}}=\underset{K\rightarrow\infty}{\lim}\frac{\overline{q}_{X}\left(\frac{e^{2}M_{K}}{K}\right)}{\left(\frac{\ln\left(K\right)+\sqrt{\ln\left(K\right)}}{\lambda}\right)^{1/\gamma}}=\underset{K\rightarrow\infty}{\lim}\frac{\overline{q}_{X}\left(\frac{e^{2}M_{K}}{K}\right)}{\overline{q}_{X}\left(\frac{1}{K}\right)}\underset{K\rightarrow\infty}{\lim}\frac{\overline{q}_{X}\left(\frac{1}{K}\right)}{\left(\frac{\ln\left(K\right)+\sqrt{\ln\left(K\right)}}{\lambda}\right)^{1/\gamma}}=1.\label{eq:25}
\end{equation}
Finally, because by definition $\frac{X_{(K-M_{K}+1)}}{X_{(K)}}\leq1$
, the limits$\underset{K\rightarrow\infty}{\lim}\frac{L_{K}}{U_{K}}=1$
and $\underset{K\rightarrow\infty}{\lim}\Pr\left(\frac{X_{(K-M_{K}+1)}}{X_{(K)}}\geq\frac{L_{K}}{U_{K}}\right)=1$
suggest that for each $\delta>0$ the limit $\underset{K\rightarrow\infty}{\lim}\Pr\left(1-\frac{X_{(K-M_{K}+1)}}{X_{(K)}}\geq\delta\right)=0$
holds.
\end{IEEEproof}
Next we prove that the non-sharing users, which are almost all of
the users, maintain max-min fairness between them.
\begin{IEEEproof}[Proof of Theorem \ref{Max-Min Fairness}]
Denote $\underline{n}=\arg\underset{n}{\min}P_{n}a_{n}X_{n,(K-M_{K}+1)}$
and $\overline{n}=\arg\underset{n}{\max}P_{n}a_{n}X_{n,(K)}$. By
combining Lemma \ref{Extreme Statistics Lemma} Part 2 and Lemma \ref{Intermediate Statistics Lemma}
Part 2 we obtain, for $L_{K}=\bar{q}_{X}\left(\frac{e^{2}M_{K}}{K}\right)$
(see Definition \ref{def:Tail Quantile}) and $U_{K}=\left(\frac{\ln\left(K\right)+\sqrt{\ln\left(K\right)}}{\lambda}\right)^{1/\gamma}$,
that
\begin{multline}
\Pr\Biggl(\frac{\underset{n}{\min}\log_{2}\left(1+\frac{P_{n}}{N_{0}}a_{n}X_{n,(K-M_{K}+1)}\right)}{\underset{n}{\max}\log_{2}\left(1+\frac{P_{n}}{N_{0}}a_{n}X_{n,(K)}\right)}\geq\frac{\log_{2}\left(1+\frac{P_{\underline{n}}}{N_{0}}a_{\underline{n}}L_{K}\right)}{\log_{2}\left(1+\frac{P_{\overline{n}}}{N_{0}}a_{\overline{n}}U_{N^{2}}\right)}\Biggr)\underset{(a)}{\geq}\\
\Pr\left(\underset{n}{\min}X_{n,(K-M_{K}+1)}\geq L_{K}\,,\,\underset{n}{\max}X_{n,(K)}\leq U_{N^{2}}\right)\underset{(b)}{\geq}\\
1-\Pr\left(\underset{n}{\min\,}X_{n,(K-M_{K}+1)}\leq L_{K}\right)-\Pr\left(\underset{n}{\max\,}X_{n,(K)}\geq U_{N^{2}}\right)\underset{K\rightarrow\infty}{\rightarrow}1\label{eq:26}
\end{multline}
where (a) follows because $\underset{n}{\min}X_{n,(K-M_{K}+1)}\geq L_{K}$
and $\underset{n}{\max}X_{n,(K)}\leq U_{N^{2}}$ imply this inequality,
and also due to the definitions of $\underline{n},\overline{n}$.
Inequality (b) is from Fr�chet inequality. By applying Proposition
\ref{Eponentially-Dominated Tail Distribution Properites} we obtain
\begin{equation}
\underset{N\rightarrow\infty}{\lim}\frac{L_{N}}{U_{N^{2}}}=\underset{N\rightarrow\infty}{\lim}\frac{\bar{q}_{X}\left(\frac{e^{2}M_{N}}{N}\right)}{\left[\frac{\ln\left(N^{2}\right)+\sqrt{\ln\left(N^{2}\right)}}{\lambda}\right]^{1/\gamma}}\underset{(a)}{=}\underset{N\rightarrow\infty}{\lim}\frac{\left[\frac{1}{\lambda}\ln\left(N^{2}\right)\right]^{1/\gamma}}{\left[\frac{\ln\left(N^{2}\right)+\sqrt{\ln\left(N^{2}\right)}}{\lambda}\right]^{1/\gamma}}\underset{N\rightarrow\infty}{\lim}\frac{\bar{q}_{X}\left(\frac{e^{2}M_{N}}{N}\right)}{\bar{q}_{X}\left(\frac{1}{N^{2}}\right)}\underset{(b)}{>}0\label{eq:27}
\end{equation}
where (a) follows from $\underset{N\rightarrow\infty}{\lim}\frac{\bar{q}_{X}\left(\frac{1}{N^{2}}\right)}{\left[\frac{1}{\lambda}\ln\left(N^{2}\right)\right]^{1/\gamma}}=1$
(Part 1 of Proposition \ref{Eponentially-Dominated Tail Distribution Properites}).
Inequality (b) follows from $\left[\frac{1}{\lambda}\ln\left(N^{2}\right)\right]^{1/\gamma}$
being the dominant term in $\left[\frac{\ln\left(N^{2}\right)+\sqrt{\ln\left(N^{2}\right)}}{\lambda}\right]^{1/\gamma}$and
from part 3 of Proposition \ref{Eponentially-Dominated Tail Distribution Properites}.
Denote $\xi=\min\left\{ 1,\frac{P_{\underline{n}}a_{\underline{n}}}{P_{\overline{n}}a_{\overline{n}}}\right\} $.
We conclude that
\begin{equation}
\underset{N\rightarrow\infty}{\lim}\frac{\log_{2}\left(1+\frac{P_{\underline{n}}}{N_{0}}a_{\underline{n}}L_{N}\right)}{\log_{2}\left(1+\frac{P_{\overline{n}}}{N_{0}}a_{\overline{n}}U_{N^{2}}\right)}\underset{(a)}{\geq}\underset{N\rightarrow\infty}{\lim}\frac{\log_{2}\left(\left(1+\frac{P_{\overline{n}}}{N_{0}}a_{\overline{n}}U_{N^{2}}\right)\xi\frac{L_{N}}{U_{N^{2}}}\right)}{\log_{2}\left(1+\frac{P_{\overline{n}}}{N_{0}}a_{\overline{n}}U_{N^{2}}\right)}=\underset{N\rightarrow\infty}{\lim}\frac{\log_{2}\left(\xi\right)+\log_{2}\left(\frac{L_{N}}{U_{N^{2}}}\right)}{\log_{2}\left(1+\frac{P_{\overline{n}}}{N_{0}}a_{\overline{n}}U_{N^{2}}\right)}+1\underset{(b)}{=}1\label{eq:28}
\end{equation}
where (a) is from $\xi\frac{L_{N}}{U_{N^{2}}}\le1$, since either
$P_{\underline{n}}a_{\underline{n}}\geq P_{\overline{n}}a_{\overline{n}}$
and $\xi=1$ or $\xi<1$. Equality (b) is from \eqref{eq:27} since
$U_{N}$ is monotonically increasing.
\end{IEEEproof}

\section{$m$-Dependent Auxiliary Lemma\label{Correlated-Fading-Channels}}
\begin{lem}
\label{lem:m-dependent bounds}Let $\left\{ X_{k}\right\} $ be a
random process of identically distributed and unbounded random variables.
Assume that $\left\{ X_{k}\right\} $ are $m$-dependent. Let $M_{K}$
be a sequence such that $\underset{K\rightarrow\infty}{\lim}\frac{M_{K}}{K}=0$.
Define $\mathcal{M}_{K}=\left\{ k\,|\,X_{k}\geq X_{(K-M_{K}+1)}\right\} $
and $\mathcal{E}\subseteq\left\{ 1,...,K\right\} $ with $E=\left|\mathcal{E}\right|.$
If $M_{K}\geq\left(m+1\right)(e+\varepsilon)\ln\left(K\right)$ for
some $\varepsilon>0$ then there exists an $K_{0}$ such that for
each $K>K_{0}$ we have
\begin{equation}
\Pr\left(X_{(K-M_{K}+1)}>\bar{q}_{X}\left(\frac{M_{K}}{e\left(K+m+1\right)}\right)\right)\leq\frac{m+1}{K^{1+\frac{\varepsilon}{3}}}\label{eq:29}
\end{equation}
and
\begin{equation}
\Pr\left(\mathcal{M}_{K}\cap\mathcal{E}=\emptyset\right)\leq\left(m+1\right)e^{-\frac{M_{K}}{e\left(m+1\right)}\frac{1-\frac{m+1}{K}}{1+\frac{m+1}{K}}}+\left(1-\frac{M_{K}}{e\left(K+m+1\right)}\right)^{\frac{E}{m+1}}.\label{eq:30-1}
\end{equation}
\end{lem}
\begin{IEEEproof}
Denote $\delta=\bar{q}_{X}\left(\frac{M_{K}}{e\left(K+m+1\right)}\right)$
and recall that $\bar{q}_{X}\left(p\right)=F_{X}^{-1}\left(1-p\right)=\min\left\{ x\,|\,F_{X}\left(x\right)\geq1-p\right\} .$
We use Fr�chet inequality to obtain
\begin{equation}
\Pr\left(k\in\mathcal{M}_{K}\right)\underset{(a)}{\geq}\Pr\Biggl(X_{(K-M_{K}+1)}\leq\delta,X_{k}>\delta\Biggr)\geq\Pr\left(X_{(K-M_{K}+1)}\leq\delta\right)+\Pr\left(X_{k}>\delta\right)-1\label{eq:45}
\end{equation}
where (a) follows because if $X_{k}\geq\delta$ and $X_{(K-M_{K}+1)}\leq\delta$
then $k\in\mathcal{M}_{K}$. We want to find a lower bound for the
first probability in \eqref{eq:45}. This is similar to the proof
of Lemma \ref{Intermediate Statistics Lemma} with reversed inequalities.
First invoke $F_{K}\left(x\right)=\sum_{k=1}^{K}I\left(X_{k}\leq x\right)$
to obtain
\begin{equation}
\Pr\left(X_{(K-M_{K}+1)}>\delta\right)=\Pr\left(\sum_{k=1}^{K}I\left(X_{k}>\delta\right)\geq M_{K}\right).\label{eq:46}
\end{equation}
We obtain
\begin{multline}
\Pr\left(\sum_{k=1}^{K}I\left(X_{k}>\delta\right)\geq M_{K}\right)\underset{(a)}{\leq}\Pr\left(\bigcup_{j=1}^{m+1}\left\{ \sum_{i=0}^{s_{j}}I\left(X_{j+\left(m+1\right)i}>\delta\right)\geq\frac{M_{K}}{m+1}\right\} \right)\underset{(b)}{\leq}\\
\left(m+1\right)\Pr\left(\sum_{i=0}^{\underset{j}{\max}s_{j}}I\left(X_{1+\left(m+1\right)i}>\delta\right)\geq\frac{M_{K}}{m+1}\right)\label{eq:48}
\end{multline}
where
\[
s_{j}=\left\{ \begin{array}{cc}
\left\lfloor \frac{K}{m+1}\right\rfloor -1 & j>K-\left\lfloor \frac{K}{m+1}\right\rfloor \left(m+1\right)\\
\left\lfloor \frac{K}{m+1}\right\rfloor  & else
\end{array}\right.
\]
and (a) follows because some inner sum must be at least $\frac{M_{K}}{m+1}$
in order for the total sum to be at least $M_{K}$, and (b) from the
union bound. We can apply a concentration upper bound on the last
term, denoting the success probability of the corresponding Bernoulli
process $Z_{i}=I\left(X_{1+\left(m+1\right)i}>\delta\right)$ by $p=1-F_{X}\left(\delta\right)=\frac{M_{K}}{e\left(K+m+1\right)}<\frac{M_{K}}{K}$.
Denote $S=\underset{j}{\max}s_{j}+1$. According to Theorem A.1.12
in \cite{Alon2015}, for all $N>0$ and $\beta>1$
\begin{equation}
\Pr\left(\sum_{i=0}^{\underset{j}{\max}s_{j}}Z_{i}\geq\beta pS\right)\leq\left(e^{\beta-1}\beta^{-\beta}\right)^{pS}\label{eq:54}
\end{equation}
where here
\begin{equation}
\beta=\frac{\frac{M_{K}}{m+1}}{\frac{M_{K}S}{e\left(K+m+1\right)}}\underset{(a)}{\geq}\frac{\frac{M_{K}}{m+1}}{\frac{M_{K}\left(\frac{K}{m+1}+1\right)}{e\left(K+m+1\right)}}=e\label{eq:55}
\end{equation}
where (a) is due $S=\underset{j}{\max}s_{j}+1\leq\frac{K}{m+1}+1$.
We obtain
\begin{equation}
\Pr\left(\sum_{i=0}^{\underset{j}{\max}s_{j}}Z_{i}\geq\frac{M_{K}}{m+1}\right)\underset{(a)}{\leq}e^{-S\frac{M_{K}}{e(K+m+1)}}\underset{(b)}{\leq}e^{-\frac{M_{K}}{e\left(m+1\right)}\frac{1-\frac{m+1}{K}}{1+\frac{m+1}{K}}}\label{eq:56}
\end{equation}
where (a) is due to \eqref{eq:54} and \eqref{eq:55} and (b) due
to $S=\underset{j}{\max}s_{j}+1\geq\frac{K}{m+1}-1$. If $M_{K}\geq\left(m+1\right)(e+\varepsilon)\ln\left(K\right)$
for some $\varepsilon>0$ then by \eqref{eq:46},\eqref{eq:48} and
\eqref{eq:56} we conclude that for large enough $K$
\begin{equation}
\Pr\left(X_{(K-M_{K}+1)}>\delta\right)\leq\left(m+1\right)e^{-\frac{M_{K}}{e\left(m+1\right)}\frac{1-\frac{m+1}{K}}{1+\frac{m+1}{K}}}\leq\left(m+1\right)\left(\frac{1}{K^{1+\frac{\varepsilon}{e}}}\right)^{\frac{1-\frac{m+1}{K}}{1+\frac{m+1}{K}}}\underset{(a)}{\leq}\frac{m+1}{K^{1+\frac{\varepsilon}{3}}}\label{eq:57}
\end{equation}
where (a) holds for large enough $K$. Using the above bound on \eqref{eq:45}
we conclude that for large enough $K$
\begin{equation}
\Pr\left(k\in\mathcal{M}_{K}\right)\underset{\left(a\right)}{\geq}\frac{M_{K}}{e\left(K+m+1\right)}-\frac{m+1}{K^{1+\frac{\varepsilon}{3}}}\geq\frac{M_{K}}{3K}\label{eq:52}
\end{equation}
where (a) uses $\Pr\left(X_{k}>\delta\right)=\frac{M_{K}}{e\left(K+m+1\right)}$
and \eqref{eq:57}. Now observe that at least $\frac{E}{m+1}$ REs
in $\mathcal{E}$ are independent. Denote their set of indices by
$\mathcal{I}$. Hence

\begin{equation}
\Pr\left(\bigcup_{k\in E}\left\{ X_{k}>\delta\right\} \right)\geq\Pr\left(\bigcup_{k\in\mathcal{I}}\left\{ X_{k}>\delta\right\} \right)\underset{(a)}{\geq}1-\left(1-\frac{M_{K}}{e\left(K+m+1\right)}\right)^{\frac{E}{m+1}}\label{eq:59}
\end{equation}
where (a) is due to $\delta=\bar{q}_{X}\left(\frac{M_{K}}{e\left(K+m+1\right)}\right)$.
We conclude that for large enough $K$
\begin{multline}
\Pr\left(\mathcal{M}_{K}\cap\mathcal{E}\neq\emptyset\right)\underset{(a)}{\geq}\Pr\Biggl(X_{(K-M_{K}+1)}\leq\delta,\bigcup_{k\in E}\left\{ X_{k}>\delta\right\} \Biggr)\underset{(b)}{\geq}\\
\Pr\left(X_{(K-M_{K}+1)}\leq\delta\right)+\Pr\left(\bigcup_{k\in E}\left\{ X_{k}>\delta\right\} \right)-1\underset{(c)}{\geq}1-\left(m+1\right)e^{-\frac{M_{K}}{e\left(m+1\right)}\frac{1-\frac{m+1}{K}}{1+\frac{m+1}{K}}}-\left(1-\frac{M_{K}}{e\left(K+m+1\right)}\right)^{\frac{E}{m+1}}\label{eq:53}
\end{multline}
where (a) follows because if $X_{k}>\delta$ and $X_{(K-M_{K}+1)}\leq\delta$,
then $k\in\mathcal{M}_{K}$. Inequality (b) is the Fr�chet inequality.
Inequality (c) follows from the bounds in \eqref{eq:57} and \eqref{eq:59}.
This means that for large enough $K$, $\Pr\left(\mathcal{M}_{K}\cap\mathcal{E}=\emptyset\right)\leq\left(m+1\right)e^{-\frac{M_{K}}{e\left(m+1\right)}\frac{1-\frac{m+1}{K}}{1+\frac{m+1}{K}}}+\left(1-\frac{M_{K}}{e\left(K+m+1\right)}\right)^{\frac{E}{m+1}}$.
\end{IEEEproof}

\section{Proof for Existence of Perfect Matchings in the User-Resource graph\label{Existence Proof Appendix}}

In this appendix we prove the existence of a perfect matching between
users and $M$-best REs in the general case of the $m$-dependent
frequency-selective channel. This existence theorem is used both for
the Naive-FSIG and the M-FSIG.

Our proof is based on the following famous theorem by Erd\H{o}s and
R�nyi \cite[Theorem 6.1, Page 83]{Erdos1964,Frieze2015}.
\begin{thm}
\label{Erdos-Renyi}Let $G_{p}$ be a balanced bipartite graph with
$N$ vertices at each side. We generate the edges of $G_{p}$ by selecting
each edge, independently, as a random Bernoulli variable with parameter
$p$. Denote the event in which $G_{p}$ has a perfect matching by
$A_{G_{p}}$. If $p=\frac{\ln N+c_{N}}{N}$ then
\begin{equation}
\underset{N\rightarrow\infty}{\lim}\Pr\left(A_{G_{p}}\right)=e^{-2e^{-c_{N}}}.\label{eq:60}
\end{equation}
\end{thm}
Our proof strategy is as follows. We define $m+1$ Erd\H{o}s\textendash R�nyi
disjoint graphs that are induced by our channel gains. We prove that
with a probability that goes to one as $N\rightarrow\infty$, the
union of all these Erd\H{o}s\textendash R�nyi graph is a subgraph
of our user-resource bipartite graph (see Subsection \ref{Bipartite Graphs})
with parameter $M$. Using the theorem above for each Erd\H{o}s\textendash R�nyi
graph separately, we know that the probability for a perfect matching
goes to 1 as $N\rightarrow\infty$ and therefore also their union
has a perfect matching. Together we conclude that the probability
our user-resource graph has a perfect matching goes to 1 as $N\rightarrow\infty$.
\begin{defn}
Define for each $i=0,...,m$ the indices $\mathcal{I}_{i}=\left\{ j\,|\,\left(j-i-1\right)\mod\left(m+1\right)=0,\,1\leq j\leq N\right\} $
(from $1+i$ to $N$ in jumps with size $m+1$). The $i$-th user-resource
core graph, denoted $G_{m,i}$, is a bipartite graph with $\left|\mathcal{I}_{i}\right|$
edges in each side ($\left\lfloor \frac{N}{m+1}\right\rfloor $ or
$\left\lfloor \frac{N}{m+1}\right\rfloor +1$). At the left side the
vertices are the users with indices $\mathcal{I}_{i}$. At the right
side the vertices are the REs with indices $\mathcal{I}_{i}$. An
edge between user $n$ and RE $k$ exists if and only if $X_{n,k}>\bar{q}_{X}\left(\frac{M}{e\left(K+m+1\right)}\right)$,
where $\bar{q}_{X}$ is the tail quantile function of $X_{n,k}$ (see
Definition \ref{def:Tail Quantile}). The variable $X_{n,k}$ can
be chosen as either $X_{n,k}=\left|h_{n,n,k}\right|$ or $X_{n,k}=-\left|h_{n,n,k}\right|$
(for obtaining the worst REs instead). Denote by $G_{m}=\bigcup_{i}G_{m,i}$
the union of all the user-resource core graphs, which has $N$ vertices
on each side.
\end{defn}
\begin{lem}
\label{ER perfect matching lemma} If $M\geq\left(m+1\right)\left(e+\varepsilon\right)\ln\left(N\right)$
, then the probability that $G_{m}$ has a perfect matching approaches
one as $N\rightarrow\infty$.
\end{lem}
\begin{IEEEproof}
Denote $X_{n,k}=\left|h_{n,n,k}\right|$. For the case of the $M$
worst REs of each user, we can use $X_{n,k}=-\left|h_{n,n,k}\right|$
instead. Since $M\geq\left(m+1\right)\left(e+\varepsilon\right)\ln\left(N\right)$,
we have for each $i$
\begin{multline}
p=\Pr\left(X_{n,k}>\bar{q}_{X}\left(\frac{M}{e\left(N+m+1\right)}\right)\right)=\frac{M}{e\left(N+m+1\right)}\geq\frac{\left(m+1\right)\left(1+\frac{\varepsilon}{e}\right)\ln N}{N+m+1}\underset{(a)}{\geq}\\
\frac{\left(1+\frac{\varepsilon}{e}\right)\left(m+1\right)\ln\left(\frac{N}{m+1}\right)}{N-m-1}\geq\frac{\left(1+\frac{\varepsilon}{e}\right)\ln\left(\left\lfloor \frac{N}{m+1}\right\rfloor \right)}{\left\lfloor \frac{N}{m+1}\right\rfloor }\label{eq:61}
\end{multline}
where (a) follows since $\frac{\ln N}{N+m+1}\geq\frac{\ln N-\ln\left(m+1\right)}{N-m-1}$,
which is equivalent to $\left(1+\frac{2m+2}{N-m-1}\right)\left(1-\frac{\ln\left(m+1\right)}{\ln N}\right)\leq1$,
holds for large enough $N$. Note that the larger core graphs with
$\left\lfloor \frac{N}{m+1}\right\rfloor +1$ edges require a less
restrictive inequality than in \eqref{eq:61}. Therefore, by substituting
$c_{N}=\frac{\varepsilon}{e}\ln\left(\left\lfloor \frac{N}{m+1}\right\rfloor \right)$
in Theorem \ref{Erdos-Renyi}, we have $\underset{N\rightarrow\infty}{\lim}\Pr\left(A_{G_{m,i}}\right)=1$
for each $i$. By the Fr�chet inequality and the fact that $m$ is
a constant with respect to $N$ we conclude that also$\underset{N\rightarrow\infty}{\lim}\Pr\left(\bigcap_{i}A_{G_{m,i}}\right)=1$.
A union of bipartite graphs must have a perfect matching if its disjoint
subgraphs (with $N$ vertices in total) all have a perfect matching.
Hence $\underset{N\rightarrow\infty}{\lim}\Pr\left(A_{G_{m}}\right)=1.$
\end{IEEEproof}
\begin{lem}
\label{subgraph lemma}Let $B_{M}$ be the bipartite user-resource
graph of Definition \ref{def:Player-Channel Graph}, where users are
connected to their $M$-best (or worst) REs. If $M\geq\left(m+1\right)\left(e+\varepsilon\right)\ln\left(N\right)$
then the probability that $G_{m}\subseteq B_{M}$ approaches one as
$N\rightarrow\infty$.
\end{lem}
\begin{IEEEproof}
Denote $X_{n,k}=\left|h_{n,n,k}\right|$ and $\delta=\bar{q}_{X}\left(\frac{M}{e\left(K+m+1\right)}\right)$
(for the case of the $M$ worst REs of each user, we use $X_{n,k}=-\left|h_{n,n,k}\right|$
instead). If $X_{n,(K-M+1)}\leq\delta$ then $X_{n,k}>\delta$ implies
that $X_{n,k}>X_{n,(K-M+1)}$ so by definition $k\in\mathcal{M}_{n}$
(user $n$ is connected to RE $k$). Hence, all the edges in $G_{m}$
must appear also in $B_{M}$ provided that $X_{n,(K-M+1)}\leq\delta$
for all $n$. From Lemma \ref{lem:m-dependent bounds} we know that
\begin{equation}
\Pr\left(X_{n,(K-M+1)}>\delta\right)\leq\frac{m+1}{N^{1+\frac{\varepsilon}{3}}}.\label{eq:62}
\end{equation}
So by the union bound we obtain
\begin{equation}
\Pr\left(\underset{n}{\max}X_{n,(K-M+1)}>\delta\right)\leq\frac{m+1}{N^{\frac{\varepsilon}{3}}}\label{eq:63}
\end{equation}
which concludes the proof.
\end{IEEEproof}
We conclude this appendix by proving the existence of a perfect matching
in our user-resource graph $B_{M}$.
\begin{thm}
Let $B_{M}$ be the bipartite user-resource graph of Definition \ref{def:Player-Channel Graph},
where users are connected to their $M$-best (or worst) REs. If $M\geq\left(m+1\right)\left(e+\varepsilon\right)\ln\left(N\right)$
then the probability that $B_{M}$ has a perfect matching approaches
one as $N\rightarrow\infty$.
\end{thm}
\begin{IEEEproof}
Denote the event in which $B_{M}$ has a perfect matching by $A_{M}$.
If $M\geq\left(m+1\right)\left(e+\varepsilon\right)\ln\left(N\right)$
then according to Lemma \ref{subgraph lemma} and Lemma \ref{ER perfect matching lemma}
we have, from Fr�chet inequality, that
\begin{equation}
\Pr\left(A_{M}\right)\geq\Pr\left(\underset{n}{\max}X_{n,(K-M+1)}\leq\delta\,,\,A_{G_{m}}\right)\geq\Pr\left(\underset{n}{\max}X_{n,(K-M+1)}\leq\delta\right)+\Pr\left(A_{G_{m}}\right)-1\label{eq:64}
\end{equation}
so $\Pr\left(A_{M}\right)\rightarrow1$ as $N\rightarrow\infty$.
\end{IEEEproof}
\bibliographystyle{IEEEtran}
\bibliography{itreferences}

\begin{thebibliography}{10}
\providecommand{\url}[1]{#1}
\csname url@samestyle\endcsname
\providecommand{\newblock}{\relax}
\providecommand{\bibinfo}[2]{#2}
\providecommand{\BIBentrySTDinterwordspacing}{\spaceskip=0pt\relax}
\providecommand{\BIBentryALTinterwordstretchfactor}{4}
\providecommand{\BIBentryALTinterwordspacing}{\spaceskip=\fontdimen2\font plus
\BIBentryALTinterwordstretchfactor\fontdimen3\font minus
  \fontdimen4\font\relax}
\providecommand{\BIBforeignlanguage}[2]{{%
\expandafter\ifx\csname l@#1\endcsname\relax
\typeout{** WARNING: IEEEtran.bst: No hyphenation pattern has been}%
\typeout{** loaded for the language `#1'. Using the pattern for}%
\typeout{** the default language instead.}%
\else
\language=\csname l@#1\endcsname
\fi
#2}}
\providecommand{\BIBdecl}{\relax}
\BIBdecl

\bibitem{Bistritz2015}
I.~Bistritz and A.~Leshem, ``Asymptotically optimal distributed channel
  allocation: a competitive game-theoretic approach,'' in \emph{Communication,
  Control, and Computing (Allerton), 2015 53nd Annual Allerton Conference on},
  2015.

\bibitem{Bistritz2017}
------, ``Game theoretic resource allocation for m-dependent channel with
  application to {OFDMA},'' in \emph{The 42nd IEEE International Conference on
  Acoustics, Speech and Signal Processing}, 2017.

\bibitem{TeSun1981}
H.~Te~Sun and K.~Kobayashi, ``A new achievable rate region for the interference
  channel,'' \emph{IEEE Transactions on Information Theory}, vol.~27, no.~1,
  pp. 49--60, 1981.

\bibitem{Etkin2008}
R.~H. Etkin, D.~N. Tse, and H.~Wang, ``{G}aussian interference channel capacity
  to within one bit,'' \emph{IEEE Transactions on Information Theory}, vol.~54,
  no.~12, pp. 5534--5562, 2008.

\bibitem{Cadambe2008}
V.~R. Cadambe and S.~A. Jafar, ``Interference alignment and degrees of freedom
  of the-user interference channel,'' \emph{IEEE Transactions on Information
  Theory}, vol.~54, no.~8, pp. 3425--3441., 2008.

\bibitem{ElAyach2013}
O.~El~Ayach, S.~W. Peters, and R.~Heath, ``The practical challenges of
  interference alignment,'' \emph{IEEE Wireless Communications}, vol.~20,
  no.~1, pp. 35--42, 2013.

\bibitem{Leshem2006}
A.~Leshem and E.~Zehavi, ``Bargaining over the interference channel,'' in
  \emph{Information Theory, 2006 IEEE International Symposium on}, 2006.

\bibitem{Leshem2008}
------, ``Cooperative game theory and the {G}aussian interference channel,''
  \emph{Selected Areas in Communications, IEEE Journal on}, vol.~26, no.~7, pp.
  1078--1088., 2008.

\bibitem{Laufer2005}
A.~Laufer and A.~Leshem, ``Distributed coordination of spectrum and the
  prisoner~ s dilemma,'' in \emph{New Frontiers in Dynamic Spectrum Access
  Networks, 2005. DySPAN 2005. 2005 First IEEE International Symposium on},
  2005.

\bibitem{Pang2008}
J.-S. Pang, G.~Scutari, F.~Facchinei, and C.~Wang, ``Distributed power
  allocation with rate constraints in gaussian parallel interference
  channels,'' \emph{IEEE Transactions on Information Theory}, vol.~54, no.~8,
  pp. 3471--3489, 2008.

\bibitem{Berry2011}
R.~Berry, D.~N. Tse \emph{et~al.}, ``{S}hannon meets {N}ash on the interference
  channel,'' \emph{IEEE Transactions on Information Theory}, vol.~57, no.~5,
  pp. 2821--2836, 2011.

\bibitem{Noam2009}
Y.~Noam and A.~Leshem, ``Iterative power pricing for distributed spectrum
  coordination in {D}{S}{L},'' \emph{IEEE Transactions on Communications},
  vol.~57, no.~4, pp. 948--953, 2009.

\bibitem{Jorswieck2013}
E.~Jorswieck and R.~Mochaourab, ``{S}hannon meets {W}alras on interference
  networks,'' in \emph{Information Theory and Applications Workshop (ITA),
  2013}, 2013.

\bibitem{Alpcan2009}
T.~Alpcan, L.~Pavel, and N.~Stefanovic, ``A control theoretic approach to
  noncooperative game design,'' in \emph{Decision and Control, 2009 held
  jointly with the 2009 28th Chinese Control Conference. CDC/CCC 2009.
  Proceedings of the 48th IEEE Conference on}, 2009.

\bibitem{Gao2014}
L.~Gao, G.~Iosifidis, J.~Huang, and L.~Tassiulas, ``Hybrid data pricing for
  network-assisted user-provided connectivity,'' in \emph{INFOCOM, 2014
  Proceedings IEEE}, 2014, pp. 682--690.

\bibitem{Koutsoupias1999}
E.~Koutsoupias and C.~Papadimitriou, ``Worst-case equilibria,'' in \emph{STACS
  99}.\hskip 1em plus 0.5em minus 0.4em\relax Springer, 1999, pp. 404--413.

\bibitem{Saraydar2002}
C.~U. Saraydar, N.~B. Mandayam, and D.~J. Goodman, ``Efficient power control
  via pricing in wireless data networks,'' \emph{IEEE Transactions on
  Communications}, vol.~50, no.~2, pp. 291--303, 2002.

\bibitem{Alpcan2009a}
T.~Alpcan and L.~Pavel, ``{N}ash equilibrium design and optimization,'' in
  \emph{Game Theory for Networks, 2009. GameNets' 09. International Conference
  on}, 2009.

\bibitem{Marden2013}
J.~R. Marden and A.~Wierman, ``Distributed welfare games,'' \emph{Operations
  Research}, vol.~61, no.~1, pp. 155--168, 2013.

\bibitem{Gopalakrishnan2011}
R.~Gopalakrishnan, J.~R. Marden, and A.~Wierman, ``An architectural view of
  game theoretic control,'' \emph{ACM SIGMETRICS Performance Evaluation
  Review}, vol.~38, no.~3, pp. 31--36, 2011.

\bibitem{Larsson2009}
E.~G. Larsson, E.~Jorswieck, J.~Lindblom, R.~Mochaourab \emph{et~al.}, ``Game
  theory and the flat-fading {G}aussian interference channel,'' \emph{IEEE
  Signal Processing Magazine}, vol.~26, no.~5, pp. 18--27, 2009.

\bibitem{Scutari2008}
G.~Scutari, D.~P. Palomar, and S.~Barbarossa, ``Competitive design of multiuser
  {MIMO} systems based on game theory: A unified view,'' \emph{Selected Areas
  in Communications, IEEE Journal on}, vol.~26, no.~7, pp. 1089--1103., 2008.

\bibitem{Leshem2009}
A.~Leshem and E.~Zehavi., ``Game theory and the frequency selective
  interference channel,'' \emph{Signal Processing Magazine, IEEE}, vol.~26,
  no.~5, pp. 28--40, 2009.

\bibitem{Tsiropoulou2016}
E.~E. Tsiropoulou, A.~Kapoukakis, and S.~Papavassiliou, ``Uplink resource
  allocation in {SC-FDMA} wireless networks: A survey and taxonomy,''
  \emph{Computer Networks}, vol.~96, pp. 1--28, 2016.

\bibitem{Yu2002}
W.~Yu, G.~Ginis, and J.~M. Cioffi, ``Distributed multiuser power control for
  digital subscriber lines,'' \emph{Selected Areas in Communications, IEEE
  Journal on}, vol.~20, no.~5, pp. 1105--1115., 2002.

\bibitem{Scutari2008a}
G.~Scutari, D.~P. Palomar, and S.~Barbarossa, ``Asynchronous iterative
  water-filling for {G}aussian frequency-selective interference channels,''
  \emph{IEEE Transactions on Information Theory}, vol.~54, no.~7, pp.
  2868--2878, 2008.

\bibitem{Scutari2009}
------, ``The {M}{I}{M}{O} iterative waterfilling algorithm,'' \emph{IEEE
  Transactions on Signal Processing}, vol.~57, no.~5, pp. 1917--1935, 2009.

\bibitem{Zhang2015}
C.~Zhang, S.~Lasaulce, and E.~V. Belmega, ``Using more channels can be
  detrimental to the global performance in interference networks,'' in
  \emph{2015 IEEE International Conference on Communication Workshop (ICCW)},
  2015.

\bibitem{Knopp1995}
R.~Knopp and P.~A. Humblet, ``Information capacity and power control in
  single-cell multiuser communications,'' in \emph{Communications, 1995. ICC'95
  Seattle,'Gateway to Globalization', 1995 IEEE International Conference on},
  1995.

\bibitem{Han2012}
Z.~Han, D.~Niyato, W.~Saad, T.~Ba{\c{s}}ar, and A.~Hj{\o}rungnes, \emph{Game
  theory in wireless and communication networks}.\hskip 1em plus 0.5em minus
  0.4em\relax Cambridge University Press, 2011.

\bibitem{Barany2007}
I.~B{\'a}r{\'a}ny, S.~Vempala, and A.~Vetta, ``{N}ash equilibria in random
  games,'' \emph{Random Structures \& Algorithms}, vol.~31, no.~4, pp.
  391--405, 2007.

\bibitem{Rinott2000}
Y.~Rinott and M.~Scarsini, ``On the number of pure strategy {N}ash equilibria
  in random games,'' \emph{Games and Economic Behavior}, vol.~33, no.~2, pp.
  274--293, 2000.

\bibitem{Cohen1998}
J.~E. Cohen, ``Cooperation and self-interest: Pareto-inefficiency of {N}ash
  equilibria in finite random games,'' \emph{Proceedings of the National
  Academy of Sciences}, vol.~95, no.~17, pp. 9724--9731, 1998.

\bibitem{Daskalakis2011}
C.~Daskalakis, A.~G. Dimakis, E.~Mossel \emph{et~al.}, ``Connectivity and
  equilibrium in random games,'' \emph{The Annals of Applied Probability},
  vol.~21, no.~3, pp. 987--1016, 2011.

\bibitem{Perlaza2013}
S.~M. Perlaza, S.~Lasaulce, and M.~Debbah, ``Equilibria of channel selection
  games in parallel multiple access channels,'' \emph{EURASIP Journal on
  Wireless Communications and Networking}, vol. 2013, no.~1, pp. 1--23, 2013.

\bibitem{Rose2001}
L.~Rose, S.~M. Perlaza, and M.~Debbah, ``On the {N}ash equilibria in
  decentralized parallel interference channels,'' in \emph{2011 IEEE
  International Conference on Communications Workshops (ICC)}, 2001.

\bibitem{Kelly1998}
F.~P. Kelly, A.~K. Maulloo, and D.~K. Tan, ``Rate control for communication
  networks: shadow prices, proportional fairness and stability,'' \emph{Journal
  of the Operational Research society}, vol.~49, no.~3, pp. 237--252, 1998.

\bibitem{Palomar2006}
D.~P. Palomar and M.~Chiang, ``A tutorial on decomposition methods for network
  utility maximization,'' \emph{IEEE Journal on Selected Areas in
  Communications}, vol.~24, no.~8, pp. 1439--1451, 2006.

\bibitem{Naparstek2013}
O.~Naparstek and A.~Leshem, ``A fast matching algorithm for asymptotically
  optimal distributed channel assignment,'' in \emph{Digital Signal Processing
  (DSP), 2013 18th International Conference on}, 2013.

\bibitem{Naparstek2014}
------, ``Fully distributed optimal channel assignment for open spectrum
  access,'' \emph{IEEE Transactions on Signal Processing}, vol.~62, no.~2, pp.
  283--294., 2014.

\bibitem{Naparstek2014a}
------, ``Expected time complexity of the auction algorithm and the push
  relabel algorithm for maximum bipartite matching on random graphs,'' in
  \emph{Random Structures \& Algorithms}.\hskip 1em plus 0.5em minus
  0.4em\relax Wiley Online Library, 2014.

\bibitem{Leshem2012}
A.~Leshem, E.~Zehavi, and Y.~Yaffe, ``Multichannel opportunistic carrier
  sensing for stable channel access control in cognitive radio systems,''
  \emph{IEEE Journal on Selected Areas in Communications}, vol.~30, no.~1, pp.
  82--95, January 2012.

\bibitem{Cohen2013}
A.~L. Cohen, Kobi and E.~Zehavi, ``Game theoretic aspects of the multi-channel
  {ALOHA} protocol in cognitive radio networks,'' \emph{Selected Areas in
  Communications, IEEE Journal on}, vol.~31, no.~11, pp. 2276--2288., 2013.

\bibitem{Chang2000}
E.~conserving routing in wireless ad-hoc networks, ``Chang, jae-hwan and
  tassiulas, leandros,'' in \emph{INFOCOM 2000. Nineteenth Annual Joint
  Conference of the IEEE Computer and Communications Societies. Proceedings.
  IEEE}, 2000, pp. 22--31.

\bibitem{Chang2004}
J.-H. Chang and L.~Tassiulas, ``Maximum lifetime routing in wireless sensor
  networks,'' \emph{IEEE/ACM Transactions on networking}, vol.~12, no.~4, pp.
  609--619, 2004.

\bibitem{Owen1995}
G.~Owen, \emph{Game Theory}.\hskip 1em plus 0.5em minus 0.4em\relax Academic
  Press, 1995.

\bibitem{Babichenko2012}
Y.~Babichenko, ``Completely uncoupled dynamics and {N}ash equilibria,''
  \emph{Games and Economic Behavior}, vol.~76, no.~1, pp. 1--14, 2012.

\bibitem{Pradelski2012}
B.~S. Pradelski and H.~P. Young, ``Learning efficient {N}ash equilibria in
  distributed systems,'' \emph{Games and Economic behavior}, vol.~75, no.~2,
  pp. 882--897, 2012.

\bibitem{Shamma2005}
J.~S. Shamma and G.~Arslan, ``Dynamic fictitious play, dynamic gradient play,
  and distributed convergence to {N}ash equilibria,'' \emph{IEEE Transactions
  on Automatic Control}, vol.~50, no.~3, pp. 312--327, 2005.

\bibitem{Rose2011}
L.~Rose, S.~Lasaulce, S.~M. Perlaza, and M.~Debbah, ``Learning equilibria with
  partial information in decentralized wireless networks,'' \emph{IEEE
  communications Magazine}, vol.~49, no.~8, pp. 136--142, 2011.

\bibitem{Brown1951}
G.~W. Brown, ``Iterative solution of games by fictitious play,'' \emph{Activity
  analysis of production and allocation}, vol.~13, no.~1, pp. 374--376, 1951.

\bibitem{Levin}
J.~Levin, ``Learning in games,''
  \url{http://web.stanford.edu/~jdlevin/Econ%20286/Learning.pdf}.

\bibitem{Perlaza2011}
S.~M. Perlaza, V.~Quintero-Florez, H.~Tembine, and S.~Lasaulce, ``On the
  convergence of fictitious play in channel selection games,'' \emph{IEEE Latin
  America Transactions}, vol.~9, no.~4, pp. 470--476, 2011.

\bibitem{Bistritz2016}
I.~Bistritz and A.~Leshem, ``Convergence of approximate best-response dynamics
  in interference games,'' in \emph{Decision and Control (CDC), 2016 IEEE 55th
  Conference on}, 2016, pp. 4433--4438.

\bibitem{Bistritz2018}
------, ``Approximate best-response dynamics in random interference games,''
  \emph{IEEE Transactions on Automatic Control}, vol.~63, no.~6, pp. 1547 --
  1548, 2018.

\bibitem{Akyildiz2002}
I.~F. Akyildiz, W.~Su, Y.~Sankarasubramaniam, and E.~Cayirci, ``Wireless sensor
  networks: a survey,'' \emph{Computer networks}, vol.~38, no.~4, pp. 393--422,
  2002.

\bibitem{David1970}
H.~A. David and H.~N. Nagaraja, \emph{Order statistics}.\hskip 1em plus 0.5em
  minus 0.4em\relax John Wiley \& Sons, Inc., 1970.

\bibitem{HallJr1948}
M.~Hall~Jr, ``Distinct representatives of subsets,'' \emph{Bulletin of the
  American Mathematical Society}, vol.~54, no.~10, pp. 922--926, 1948.

\bibitem{LTE2009}
E.~LTE, ``Evolved universal terrestrial radio access ({E-UTRA}); base station
  ({BS}) radio transmission and reception,'' \emph{ETSI TS}, vol. 136, no. 104,
  p.~V8, 2009.

\bibitem{Hoeffding1963}
W.~Hoeffding, ``Probability inequalities for sums of bounded random
  variables,'' \emph{Journal of the American statistical association}, vol.~58,
  no. 301, pp. 13--30, 1963.

\bibitem{Cover2012}
T.~M. Cover and J.~A. Thomas, \emph{Elements of information theory}.\hskip 1em
  plus 0.5em minus 0.4em\relax John Wiley \& Sons, 2012.

\bibitem{Alon2015}
N.~Alon and J.~H. Spencer, \emph{The probabilistic method}.\hskip 1em plus
  0.5em minus 0.4em\relax John Wiley \& Sons, 2015.

\bibitem{Erdos1964}
P.~Erdos and A.~Renyi, ``On random matrices,'' \emph{Magyar Tud. Akad. Mat.
  Kutat{\'o} Int. K{\"o}zl}, vol.~8, no. 455-461, p. 1964, 1964.

\bibitem{Frieze2015}
A.~Frieze and M.~Karo{\'n}ski, \emph{Introduction to random graphs}.\hskip 1em
  plus 0.5em minus 0.4em\relax Cambridge University Press, 2015.

\end{thebibliography}

\end{document}